\documentclass[sigconf]{acmart}

\usepackage{microtype}
\usepackage{graphicx}
\usepackage{subfigure}
\usepackage{booktabs} 
\usepackage{dcolumn}
\usepackage{placeins}

\setcopyright{rightsretained}
\usepackage{hyperref}
\usepackage{dirtytalk}

\acmConference[Preprint]{Preprint}{February 2019}

\usepackage[english]{babel}
\usepackage[utf8]{inputenc}
\usepackage{amsmath}
\usepackage{amssymb}
\usepackage[colorinlistoftodos]{todonotes}
\usepackage{eurosym}
\usepackage{hyperref}
\usepackage{natbib}
\usepackage{mathtools}
\usepackage{bbm}
\usepackage{nicefrac}

\DeclareMathOperator{\arccosh}{arccosh}

\newcommand{\argmin}{\operatornamewithlimits{argmin}}

\renewcommand{\vec}[1]{\mathbf{#1}}%

\begin{document}

\title{Scalable Hyperbolic Recommender Systems}

\author{Benjamin Paul Chamberlain}
\orcid{0000-0003-4331-7676}
\affiliation{%
  \institution{ASOS.com}
  \city{London}
  \country{UK}
}
 \email{ben.chamberlain@asos.com}

\author{Stephen R. Hardwick}
\affiliation{%
  \institution{ASOS.com}
  \city{London}
  \country{UK}
}

\author{David R. Wardrope}
\affiliation{%
  \institution{University College London}
  \city{London}
  \country{UK}
}

\author{Fabon Dzogang}
\affiliation{%
 \institution{ASOS.com}
 \city{London}
 \country{UK}
 }

\author{Fabio Daolio}
\affiliation{%
 \institution{ASOS.com}
 \city{London}
 \country{UK}
 }
 
 \author{Saúl Vargas}
\affiliation{%
 \institution{ASOS.com}
 \city{London}
 \country{UK}
 }


\renewcommand{\shortauthors}{Chamberlain et al.}

\begin{abstract}
We present a large scale hyperbolic recommender system. We discuss why hyperbolic geometry is a more suitable underlying geometry for many recommendation systems and cover the fundamental milestones and insights that we have gained from its development. In doing so, we demonstrate the viability of hyperbolic geometry for recommender systems, showing that they significantly outperform Euclidean models on datasets with the properties of complex networks. Key to the success of our approach are the novel choice of underlying hyperbolic model and the use of the Einstein midpoint to define an asymmetric recommender system in hyperbolic space. These choices allow us to scale to millions of users and hundreds of thousands of items.
\end{abstract}

\maketitle

\section{Introduction}
\label{sec:intro}

Hyperbolic geometry has recently been identified as a powerful tool for neural network based representation learning~\cite{Nickel2017, Chamberlain2017e}. Hyperbolic space is negatively curved, making it better suited than flat Euclidean geometry for representing relationships between objects that are organized hierarchically~\cite{Nickel2017, Nickel2018} or that can be described by graphs taking the form of complex networks~\cite{Krioukov}.

\begin{figure}
    \centering
    \vspace{3mm}
    \includegraphics[width=0.235\textwidth]{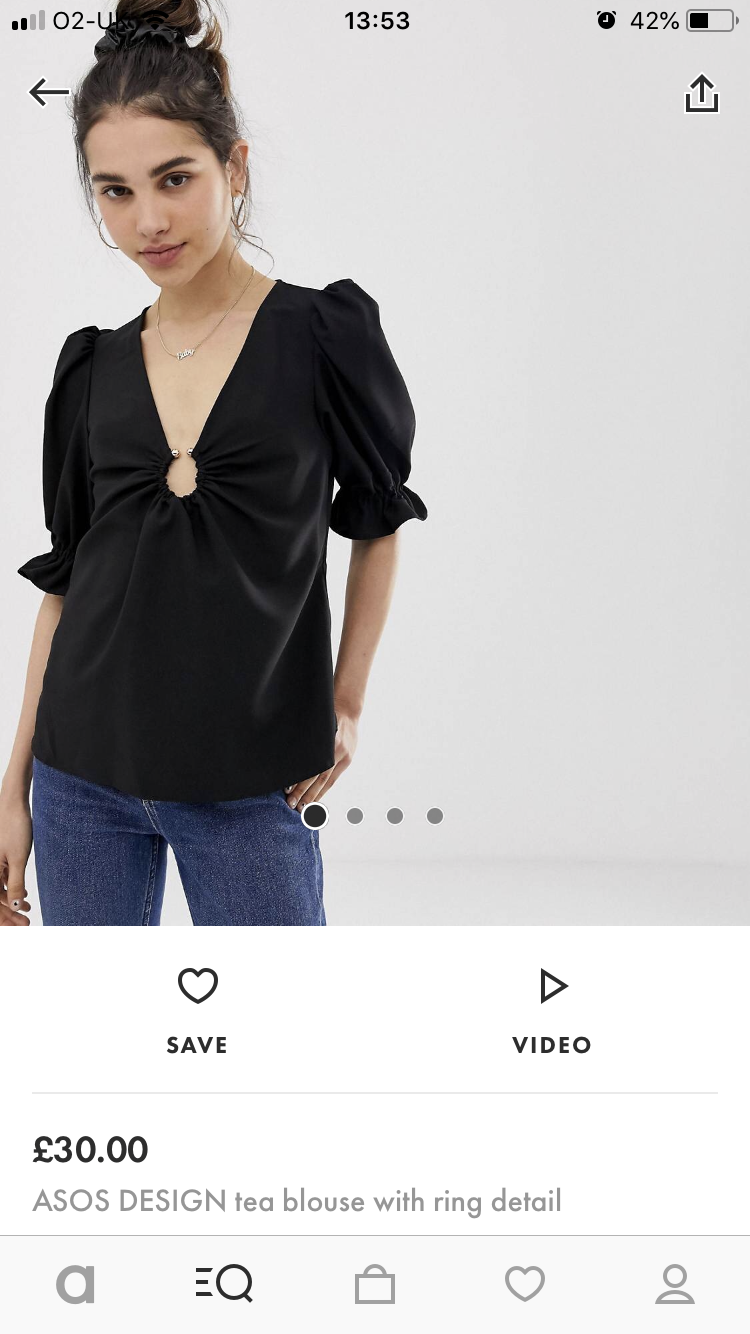}
    \includegraphics[width=0.235\textwidth]{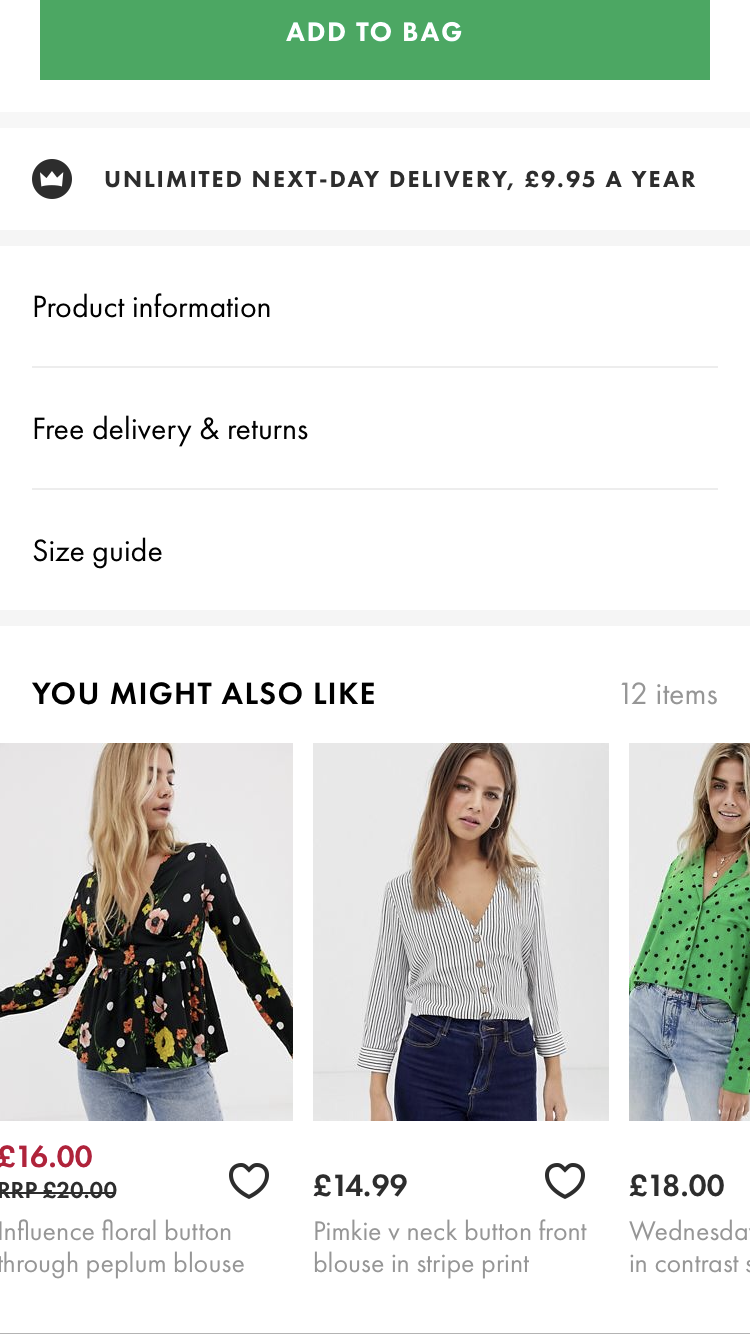}
    \caption{ASOS You Might Also Like recommendations.}
    \vspace{-5mm}
    \label{fig:recs_YMAL_example}
\end{figure}

Recommender systems are a pervasive technology, providing a major  source  of  revenue  and user satisfaction in customer facing digital businesses~\cite{Gomez-Uribe2015}. They are particularly important in e-commerce, where due to large catalogue sizes, customers are often unaware of the full extent of available products~\cite{Cardoso2018}. ASOS.com is a UK based online clothing retailer with 18.4m active customers and a live catalogue of $87,000$ products as of December 2018. Products are sold through multiple international websites and apps using a number of recommender systems. These include (1) personalised 'My Edit' recommendations that appear in the app and in emails, (2) outfit completion recommendations (3) 'You Might Also Like' recommendations that suggest alternative products, shown in Figure~\ref{fig:recs_YMAL_example} and (4) out of stock alternative recommendations. All are implicitly Euclidean, neural network based recommenders, and we believe that each of them could be improved by the use of hyperbolic geometry.

Key to the success of a recommender system is the accurate representation of user preferences and item characteristics. Matrix factorization~\cite{sarwar2000application,Hu2008} is one of most common approaches for this task. In its basic form, this technique represents user-item interactions in the form of a matrix. A factorization into two low-rank matrices, representing users and items respectively, is then computed so that it approximates the original interaction matrix. The result is a compact Euclidean vector representation for every user and every item that is useful for recommendation purposes, i.e. to estimate the interaction likelihood of unobserved pairs of users and items.

The user-item interaction matrix can be treated as the adjacency matrix of a random, undirected, bipartite graph, where edges exist between nodes representing users and items. In this paradigm user and item representations are learned by embedding nodes of the graph rather than minimising the matrix reconstruction error.
The simplest type of random graph is the Erd\H{o}s-Reny\'i or completely random graph~\cite{Erdos1960}. In this model, an edge between any two nodes is generated independently with constant probability. If user-item interaction graphs were completely random, recommendation sytems based on them would be impossible. Instead, these graphs exhibit clustering produced by similar users interacting with similar items. In addition, power law degree distributions and small world effects are also common as the rich-get-richer  effect of preferential attachment causes some users and items to have orders of magnitude more interactions than the median~\cite{barabasi1999emergence}.
Graphs of this form are known as complex networks~\cite{Newman2003}.

Completely random graphs are associated with an underlying Euclidean geometry, but the heterogenous topology of complex networks implies that the underlying geometry is hyperbolic~\cite{Krioukov}.
There are two reasons why embedding complex networks in hyperbolic geometry can be expected to perform better than Euclidean geometry. The first is that hyperbolic geometry is the continuous analogue of tree graphs~\cite{Gromov} and many complex networks display a core-periphery hierarchy and tree-like structures~\cite{adcock2013tree}. The second property is that power-law degree distributions appear naturally when points are randomly sampled in hyperbolic space and connected as an inverse function of their distance~\cite{Krioukov}.

In our hyperbolic recommender system, the Euclidean vector representations of users and items are replaced with points in hyperbolic space. As hyperbolic space is curved, the standard optimisation tools do not work and the machinery of Riemannian gradient descent must be employed~\cite{Bonnabel2013}. The additional complexity of Riemannian gradient descent is one of the major challenges to producing large-scale hyperbolic recommender systems. We mitigate this through two major innovations: (1) we carefully choose a model of hyperbolic space that permits exact gradient descent and overcomes problems with numerical instability and (2) we do not explicitly represent customers, instead implicitly representing them though the hyperbolic Einstein midpoint of their interaction histories. By doing so, our hyperbolic recommender system is able to scale to the full ASOS dataset of 18.4 million customers and one million products\footnote{87k products are live at any one time, but products are short lived and so recommendations are trained on a history containing roughly 1m products}.

We make the following contributions:
\begin{enumerate}
    \item We justify the use of hyperbolic representations for neural recommender systems through an analogy with complex networks
    \item We demonstrate that hyperbolic recommender systems can significantly outperforms Euclidean equivalents by between 2 and 14\%
    \item We develop an asymmetric hyperbolic recommender system that scales to millions of users.
\end{enumerate}

\section{Related Work}


The original work connecting hyperbolic space with complex networks was~\cite{Krioukov} and many scale-free networks such as the internet~\cite{Shavitt2008, Boguna2010} or academic citations~\cite{Clough2015a,Clough2016} have been shown to be well described by hyperbolic geometry. Hyperbolic graph embeddings were applied successfully to the problem of greedy message routing in~\cite{Kleinberg2007, Cvetkovski2009} and general graphs in low-dimensional hyperbolic space were addressed by~\cite{Blasius2016}. 

Hyperbolic geometry was introduced into the embedding layers of neural networks in \cite{Nickel2017, Chamberlain2017e} who used the Poincar\'e ball model. \cite{DeSa2018} analysed the trade-offs in numerical precision and embedding size in these different approaches and~\cite{Ganea2018} extended these models to include undirected graphs. \cite{Nickel2018} and \cite{Wilson2018} showed that the Lorentzian (or hyperboloid) model of hyperbolic space can be used to write simpler and more efficient optimisers than the Poincar\'e ball. Several works have used shallow hyperbolic neural networks to model language~\cite{Dhingra2018, Leimeister2018, Tifrea2019}. Neural networks built on Cartesian products of isotropic spaces that include hyperbolic spaces have been developed in \cite{Gu2019} and adaptive optimisers for such spaces appear in~\cite{Becigneul2019}.

Deep Hyperbolic neural networks were originally proposed in \cite{Ganea2018a} who used the formalism of M\"obius gyrovector spaces to generalise the most common Euclidean vector operations. \cite{Gulcehre2019} apply a similar approach to develop deep hyperbolic attention networks. 

\section{Background}

Hyperbolic geometry is an involved subject and comprehensive introductions appear in many textbooks (e.g. \cite{Cannon1997}). In this section we include only the material necessary for the remainder of the paper. Hyperbolic space is a homogeneous, isotropic Riemann space. A Riemann manifold $M$ is a smooth differentiable manifold. Each point on the manifold is associated with a locally Euclidean tangent space $T_\vec{x}M$. The manifold is equipped with a Riemann metric $g$ that specifies a smoothly varying positive definite inner product on $T_\vec{x}M$ at each point $\vec{x} \in M$. The shortest distance between points is not a straight line, but a geodesic curve with a distance defined by the metric tensor $g$. The map between the tangent space and the manifold is called the exponential map $f:T_\vec{x}M \to M$. As hyperbolic space can not be isometrically embedded in Euclidean space, five different  models that sit inside a Euclidean ambient space are commonly used as representations.

\subsection{Models of Hyperbolic Space}
There are multiple models of hyperbolic space because different approaches preserve some properties of the underlying space, but distort others. Each ($n$-dimensional) model has its own metric, geodesics and isometries and occupies a different subset of the ambient space $\mathbb{R}^{n+1}$. The models are all connected by simple projective maps and the most relevant for this work are the hyperboloid, and the Klein and Poincar\'e balls. As points in the models of hyperbolic space are not closed under multiplication and addition, they are not vectors in the mathematical sense. We denote them in bold font to indicate that they are one dimensional arrays.


\subsubsection{Poincar\'e Ball Model}

Much of the existing work on hyperbolic neural networks uses the Poincar\'e ball model~\cite{Nickel2017, Chamberlain2017e, Ganea2018, Ganea2018a}. It is conceptually the most simple model and our preferred choice for low dimensional visualisations of embeddings. However, gradient descent in the Poincar\'e ball is computationally complex. The Poincar\'e $n$-ball models the infinite $n$-dimensional hyperbolic space $\mathbb{H}^n$ as the interior of the unit ball. The metric tensor is
\begin{align}
g_{i,j} = \frac{4\delta_{i,j}}{(1-\left \| \mathbf{x} \right \|^2)^2},
\label{eq:poincare_metric_tensor}
\end{align}

\noindent where $\vec{x}$ is a generic point and $\delta_{i,j}$ is the kroneker delta. It is a function only of the Euclidean distance to the origin. Hyperbolic distances from the origin grow exponentially with Euclidean distance, reaching infinity at the boundary. As the metric is a point-by-point scaling of the Euclidean metric, the model is conformal.

The hyperbolic distance between Euclidean points $\vec{u}$ and $\vec{v}$ is 
\begin{align}
d(\mathbf{u},\mathbf{v}) = \arccosh (1+2\frac{\left \| \mathbf{u} - \mathbf{v} \right \|^2}{(1-\left \| \mathbf{u} \right \|^2)(1-\left \| \mathbf{v} \right \|^2)}).
\end{align}

Gradient descent optimisation within the Poincar\'e ball is challenging because the ball is bounded. Strategies to manage this problem include moving points that escape the ball back inside by a small margin~\cite{Nickel2017} or carefully managing numerical precision and other model parameters~\cite{DeSa2018}.




\subsubsection{The Klein Model}

The Klein model affords the most computational efficient calculation of the Einstein midpoint, which is used to represent user vectors as the aggregate of the item vectors. The model consists of the set of points 
\begin{equation}
    \mathbb{K}^n = \{\left(x_1,\vec{x}\right) = \left(1,\vec{x}\right) \in\mathbb{R}^{n+1} \, : \, \lVert \vec{x} \rVert < 1\}.
\end{equation}
The projection of points between the hyperboloid model and the Klein model are given by
\begin{equation}
    \mathbf{\Pi}_{\mathbb{H}\to\mathbb{K}}(x_i) = \frac{x_i}{x_1},
\end{equation}
while the inverse projection is 
\begin{equation}
\mathbf{\Pi}_{\mathbb{K}\to\mathbb{H}}(\vec{x}) = \frac{\left(1, \mathbf{x}\right)}{\sqrt{1-\lVert \vec{x} \rVert^2}}.
\end{equation}

\subsubsection{The Hyperboloid Model}

Unlike the Poincar\'e or Klein balls, the hyperboloid model is unbounded. We use the hyperboloid model as it offers efficient, closed form Riemannian Stochastic Gradient Descent (RSGD). The set of points form the upper sheet of an $n$-dimensional hyperboloid embedded in an $(n+1)$-dimensional ambient Minkowski space equipped with the following metric tensor: 
\begin{align}
g &=
\begin{bmatrix}
-1 & 0 & 0 & \dots \\ 
0 & 1 &  & \\ 
0 &  & 1 & \\ 
\vdots &  &  & \ddots  
\end{bmatrix}.
\end{align}

\noindent The inner product in Minkowski space resulting from the application of this metric tensor is
\begin{equation}
    \left<\vec{u},\vec{v}\right>_{\mathbb{H}} = -u_1v_1 + \sum_{i}^{n+1} u_iv_i.
\end{equation}

\noindent The hyperboloid can be defined as the set of points
\begin{equation}
   \{\vec{x} \in\mathbb{R}^{n+1}\quad:\quad\left<\vec{u},\vec{v}\right>_{\mathbb{H}} = -1, x_1 > 0\},
\end{equation}

\noindent where the hyperbolic distance between points $\vec{u}$ and $\vec{v}$ is defined as
\begin{equation}
    d(\vec{u}, \vec{v}) = \arccosh\left(-\left<\vec{u},\vec{v}\right>_{\mathbb{H}}\right).
\end{equation}

\noindent The tangent space $T_{\vec{x}}\mathbb{H}^n$ to a point $\vec{x}\in\mathbb{H}$, is the set of points, $\vec{v}$ satisfying
\begin{equation}
    T_{\vec{x}}\mathbb{H}^n = \{\vec{v} \, : \, \left<\vec{v},\vec{x}\right>_{\mathbb{H}} = 0\}.
\end{equation}

\noindent The projection from ambient space to tangent space is defined as
\begin{equation}
    \Pi_{\vec{x}}(\vec{v}) = \vec{v} + \left<\vec{x},\vec{v}\right>_{\mathbb{H}} \vec{x}.
\end{equation}

\noindent Finally, the exponential map from the tangent space to the hyperboloid is defined as

\begin{equation}
    \mathrm{Exp}_{\vec{x}}(\vec{v}) = \cosh\left(\|\vec{v}\|_{\mathbb{H}}\right)\vec{x} + \sinh\left(\|\vec{v}\|_{\mathbb{H}}\right)\frac{\vec{v}}{\|\vec{v}\|_{\mathbb{H}}},  
\end{equation}
where $\|\vec{v}\|_{\mathbb{H}} = \sqrt{\left<\vec{v},\vec{v}\right>_{\mathbb{H}}}$.

\section{Why Hyperbolic Geometry?}



\begin{figure}[ht]
\vskip -0.2in
\begin{center}
\includegraphics[width=0.48\columnwidth,trim={0.5cm 2.0cm 0.5cm 2.0cm},clip]{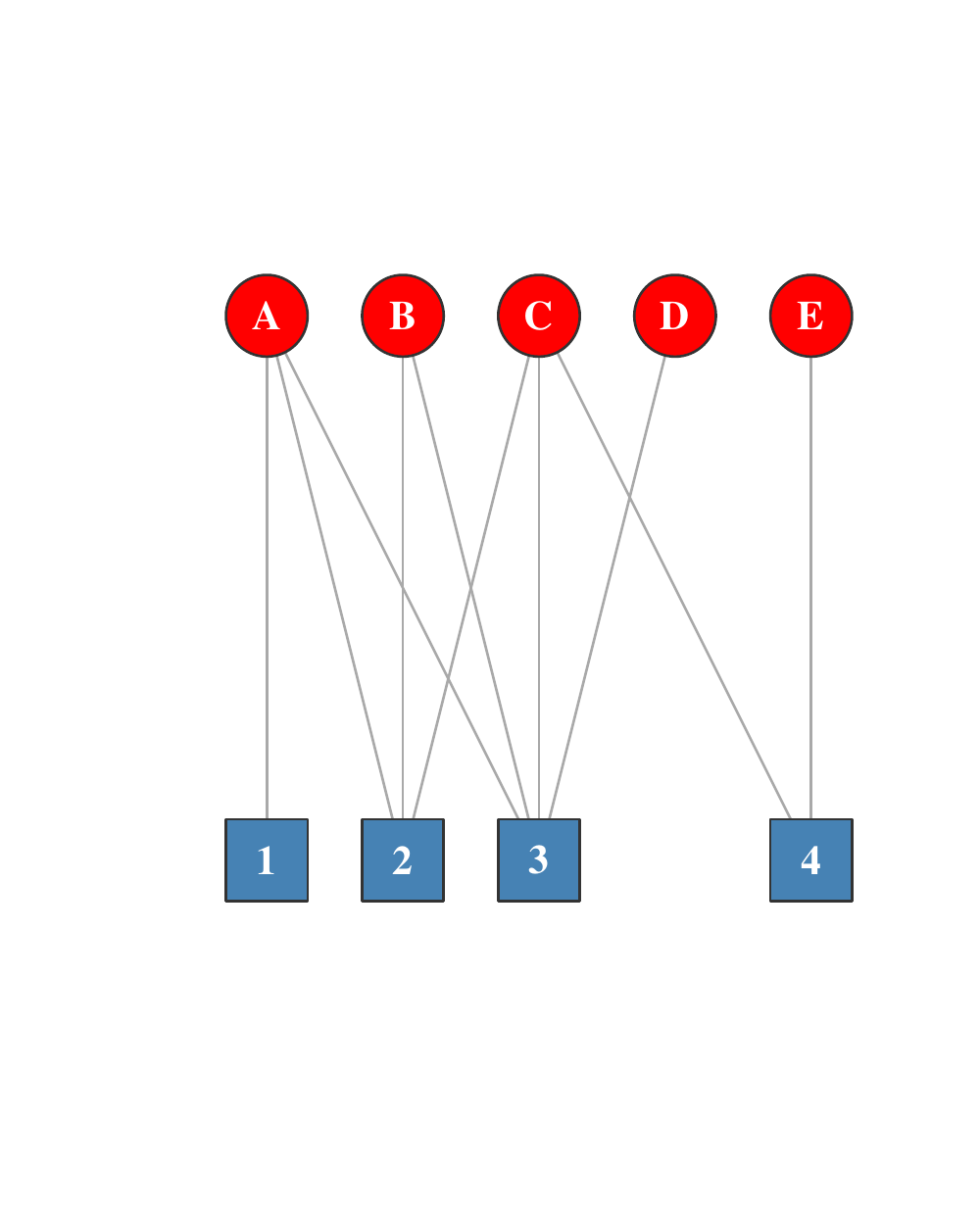}
\includegraphics[width=0.48\columnwidth,trim={0.5cm 2.0cm 0.5cm 2.0cm},clip]{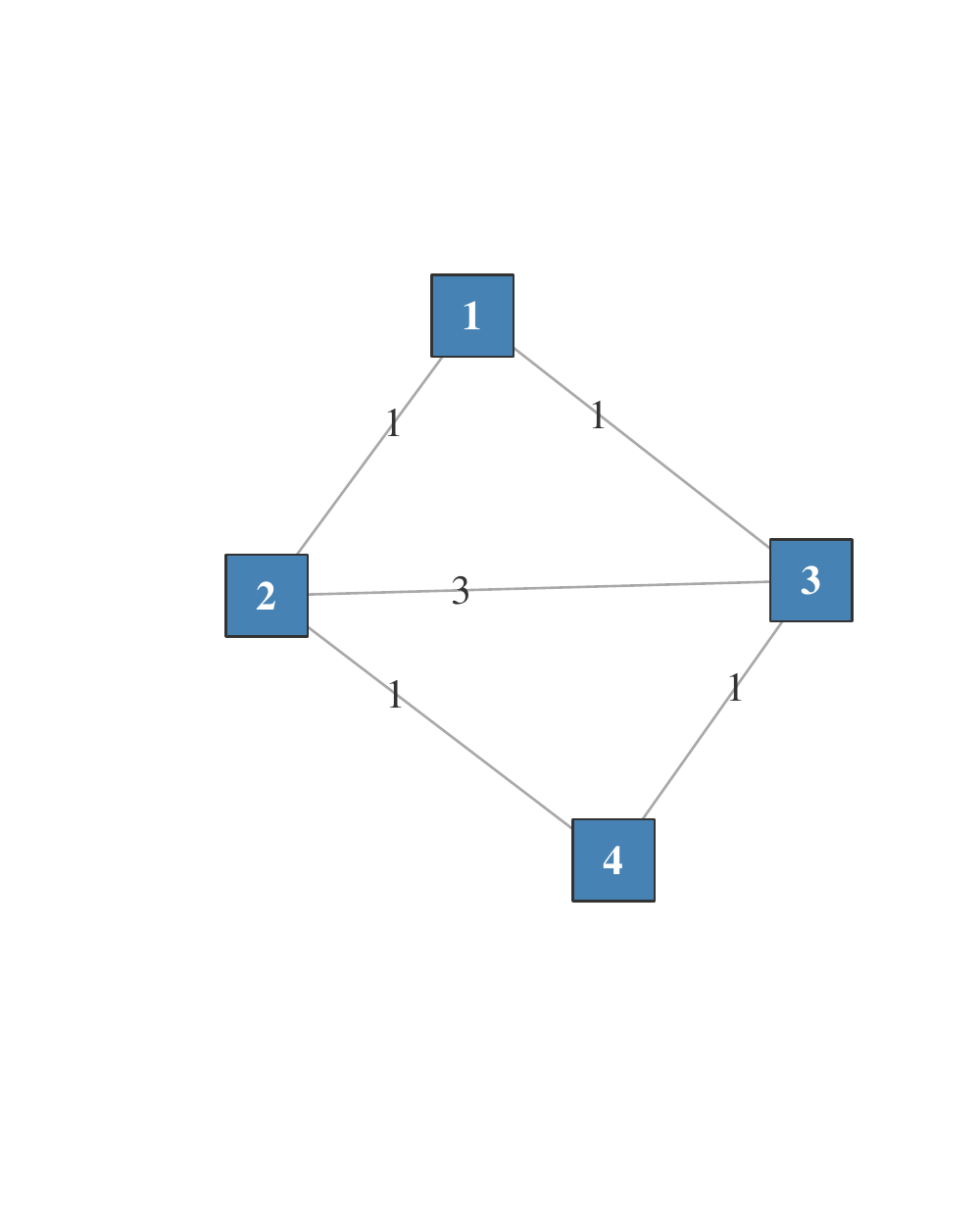}
\end{center}
\vskip -0.2in
\caption{Left: simulated user-item bipartite graph. Right: one-mode projection onto the set of items. Note that each item receives a different number of user interactions in the bipartite graph (bottom \emph{degree}), whereas the one-mode projection has a more regular structure. However, heterogeneity is partly retained in the edge weights, which account for the number of co-purchases (\emph{weighted degree}).}
\label{fig:bip}
\end{figure}

There is an intimate connection between complex networks, hyperbolic geometry and recommender systems. 
In recommender systems, the underlying graph is a two-mode, or bipartite, graph that connects users and items with an edge any time a user interacts with an item. Bipartite graphs can be projected into single-mode graphs as depicted in Figure~\ref{fig:bip} by using shared neighbour counts (or many other metrics) to represent the similarity between any pair of nodes of the same type. As such, bipartite graphs can be seen as the generative model for many complex networks~\cite{guillaume2006bipartite} e.g., the item similarity graph, is the one-mode projection of the user-item bipartite graph onto the set of items. This connection is even more explicit if we consider that, on the one hand, algorithms based on bipartite projections  have been devised to produce personalised recommendations~\cite{zhou2007bipartite}, while on the other hand, link prediction for graphs can be achieved via matrix factorisation~\cite{menon2011link}.
The topology of user-item networks and their projections has been widely studied and shown to exhibit the properties of complex networks (e.g. \cite{cano2006topology}). However, The exact influence of the underlying network structure on the performance of recommender systems remains an open question~\cite{zanin2009preferential, guo2010clustering}. 




The link between hyperbolic geometry and complex networks was established in the  seminal paper by~\cite{Krioukov} who show that 'hyperbolic geometry naturally emerges from network heterogeneity in the same way that network heterogeneity emerges from hyperbolic geometry'. If nodes are laid out uniformly at random in hyperbolic space and connected randomly as an inverse function of distance, then a complex network is obtained. Conversely, if the nodes of a complex network are treated as points in a latent metric space, where connections are more likely to form between closer nodes, then the network heterogeneous topology implies a latent hyperbolic geometry.
A similar approach has recently been applied by the same authors to characterise bipartite graphs~\cite{kitsak2017latent}.

In table~\ref{tab:stats-nets}, we report the basic statistics of the bipartite networks underlying the recommendation datasets under study. We argue that, given the complex nature of these networks, a hyperbolic space is better suited to embed them than a Euclidean one. Finally, we note that it would be a remarkable coincidence, given the large number of possibilities, if Euclidean geometry were both the \textit{only} geometry that practitioners had tried \emph{and} the optimal geometry for these problems.

\begin{table*}[!htbp] \centering 
\setlength\tabcolsep{0pt}
  \caption{Statistics of bipartite user-item graphs, and power-law fit of item degree distributions. density: proportion of interactions,  $\Bar{k}_{\text{item}}$: avg n. of interactions per item, $\hat{\gamma}$: estimated exponent of the maximum-likelihood power-law fit, KS dist: Kolmogorov-Smirnov test statistic for the distance between the data and the fitted power-law, and $p$-value of the test. Small $p$-values reject the hypothesis that the data could have been drawn from the fitted power-law distribution. The number of customers and products in the ASOS dataset are omitted for commercial reasons.}
  \label{tab:stats-nets}
  \vskip -0.1in
\begin{tabular}{D{.}{.}{-4} D{.}{.}{-4}@{\hskip 15pt\extracolsep\fill}D{.}{.}{-4} D{.}{.}{-4}@{\hskip -25pt\extracolsep\fill}D{.}{.}{-4} D{.}{.}{-4} D{.}{.}{-4} D{.}{.}{-4} D{.}{.}{-4}}
\toprule
\multicolumn{1}{l}{Data set} & & \multicolumn{1}{l}{$N_{\text{user}}$} & \multicolumn{1}{l}{$N_{\text{item}}$} & \multicolumn{1}{r}{density} & \multicolumn{1}{c}{$\Bar{k}_{\text{item}}$} & \multicolumn{1}{c}{$\hat{\gamma}$} & \multicolumn{1}{r}{KS test} & \multicolumn{1}{r}{$p$-value} \\ 
\midrule
\multicolumn{1}{l}{Automotive} & & 1,211 & 24,985 & 0.0011 & 1.3675 & 2.9767 & 0.0033 & 0.9424 \\ 
\multicolumn{1}{l}{Cell Phones and Accessories} & & 1,141 & 17,894 & 0.0016 & 1.8378 & 2.4517 & 0.0099 & 0.0611 \\ 
\multicolumn{1}{l}{Clothing Shoes and Jewelry} & & 7,917 & 165,654 & 0.0002 & 1.4241 & 2.8558 & 0.0010 & 0.9973 \\ 
\multicolumn{1}{l}{Musical Instruments} & & 471 & 11,956 & 0.0029 & 1.3801 & 3.2460 & 0.0030 & 1 \\ 
\multicolumn{1}{l}{Patio Lawn and Garden} & & 374 & 6,926 & 0.0041 & 1.5452 & 2.7184 & 0.0041 & 0.9999 \\ 
\multicolumn{1}{l}{Sports and Outdoors} & & 3,740 & 53,184 & 0.0006 & 2.1269 & 2.2874 & 0.0065 & 0.0214 \\ 
\multicolumn{1}{l}{Tools and Home Improvement} & & 2,047 & 34,422 & 0.0009 & 1.8646 & 2.3542 & 0.0157 & 0.0000\\ 
\multicolumn{1}{l}{Toys and Games} & & 3,143 & 60,361 & 0.0006 & 1.8439 & 2.4522 & 0.0060 & 0.0260 \\ 
\midrule
\multicolumn{1}{l}{MovieLens 100K} & & 942 & 1,447 & 0.0406 & 38.2688 &  5.4731 & 0.0634 &  0.9959 \\
\multicolumn{1}{l}{MovieLens 20M} & & 137,765 & 20,720 & 0.0035 & 482.3648 & 3.3181 & 0.0510 & 0.8825 \\
\midrule
\multicolumn{1}{l}{ASOS.com Menswear} & &  O(10e6) & 132,399 & 0.0000 & 198.1043 & 2.5748 & 0.0129 & 0.1286 \\
\bottomrule 
\end{tabular} 
\end{table*}

\section{Hyperbolic Recommender System}

Here we outline the overall design and individual components of our hyperbolic recommendation system, before describing each element and our detailed design choices. 

\begin{figure}
    \centering
    \includegraphics[width=0.49\textwidth, trim=5mm 20mm 5mm 8mm, clip]{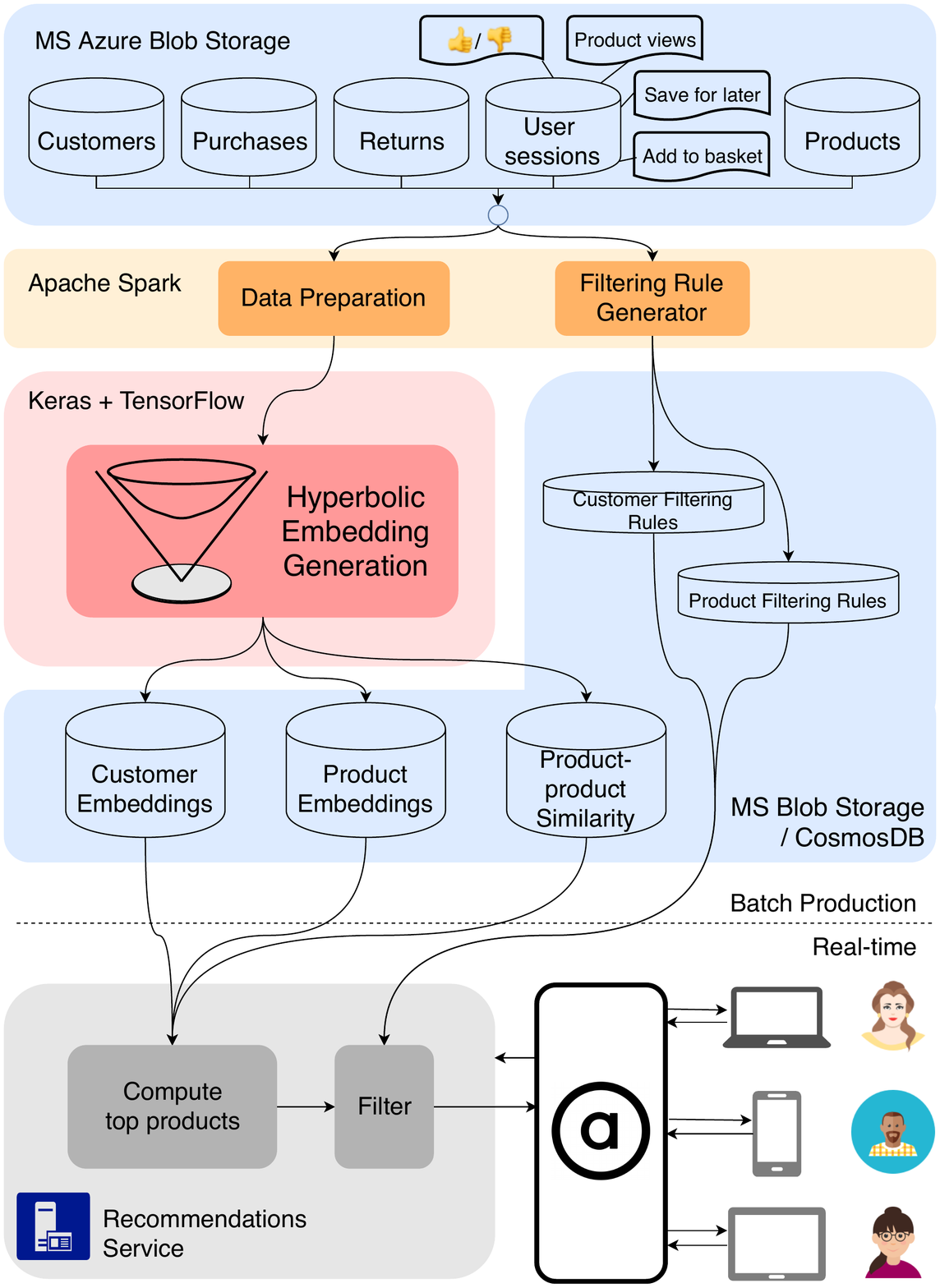}
    \caption{High level overview of the ASOS recommender system. The same infrastructure underlies multiple recommendation systems providing various user experiences.}
    \label{fig:recs_system_architecture}
\end{figure}

The recommender system is shown in Figure~\ref{fig:recs_system_architecture}. Raw data relating to customer interactions with products is stored in Microsoft Blob Storage and preprocessed into labelled customer interaction histories using Apache Spark. Hyperbolic representation learning is in  Keras~\cite{chollet2015keras} with the TensorFlow~\cite{tensorflow2015-whitepaper} back-end. The learned representations are made available to a real-time recommendation service using Cosmos DB from where they are presented to customers on web or app clients.

At a high level, our recommendation algorithm is a neural network based recommender with a learning to rank loss that represents users and items, not as Euclidean vectors, but as points in hyperbolic space. It is trained on labelled customer-product interaction histories $I_u$, where the label is the next purchased product.
As the ASOS dataset is highly asymmetric, having an order of magnitude more users than items, we do not explicitly represent users. Instead they are implicitly represented through an aggregate of the representations of the items they interact with~\cite{Cardoso2018}. For the ASOS dataset, using an asymmetric approach reduces the number of model parameters by a factor of 20 and has the additional benefit that dynamic user representations allow users' interests to change over time and with context.

Given this outline, our implementation contains four major components:
\begin{enumerate}
    \item A loss function: A ranking function to optimise
    \item A metric: Used to define item-item or user-item similarity 
    \item An optimiser: e.g. Stochastic Riemannian gradient on the hyperboloid
    \item An aggregator: To combine item representations into a user representation
\end{enumerate}

We investigated several possible approaches for each component and these are detailed in the remainder of the section.

\subsection{Loss Function}

The baseline model for the hyperbolic recommender system is Bayesian Personalized Ranking (BPR)~\cite{Rendle2009}. The BPR framework uses a triplet loss $(u, i, j)$ where $u$ indexes a user, $i$ indexes an item that they interact with and $j$ indexes a negative sample. The parameters $\Theta$, which constitute the embedding vectors are found through
\begin{align}
    \argmin_\Theta \sum_{(u,i,j)\in D} - \ln \sigma \{s_{u,i}-s_{u,j}\}+\lambda(\lVert \Theta \rVert^2),
\end{align}
where (u,i,j) sums over all pairs of positive and negative items associated with each user and $s_{u,i}$ is given by
\begin{align}
    s_{u,i} &=f(d_p(\vec{v}_u,\vec{v}_i)), 
\end{align}
where $\vec{v}_u$ and $\vec{v}_i$ are vectors representing user, $u$ and item $i$ respectively.
We acknowledge the existence of a preprint by \cite{Vinh2018} that addresses the recommendation problem in hyperbolic space using BPR. Their approach is symmetric and mirrors the setup from \cite{Nickel2017} for optimisation in the Poincar\'e ball. While \cite{Vinh2018} claim better performance for their hyperbolic recommender system than a range of Euclidean baseline models, many of the performance metrics quoted for these baseline models are worse than random and the performance of their hyperbolic systems also appears to be lower than the standard naive baseline of recommending items based on their popularity (number of interactions) in the historic data.



An alternative to BPR is the Weighted Margin-Rank Batch (WMRB) loss \cite{Liu2017WMRB:Approach}, that first approximates the rank $r$ using a set of negative samples $N$.
\begin{align}
    \mathrm{rank}(i,j) &\approx r_{i,j} = \sum_{k \in N} \lvert \epsilon + d(\vec{u}_i,\vec{v}_j) -d(\vec{u}_i,\vec{v}_k) \rvert_+ 
    \end{align}
where $d(\vec{u}_i,\vec{v}_{j,k})$ is the distance between $\vec{u}_i$ and $\vec{v}_{j,k}$, $\lvert \cdot \rvert_+ = \max(0, \cdot) $ is the ReLU activation and $\epsilon \in \mathbb{R}_+$ is a slack parameter such that terms only contribute to the loss if $d(\vec{u}_i,\vec{v}_j) + \epsilon >  d(\vec{u}_i,\vec{v}_k)$. 

WMRB calculates a pseudo-ranking for the positive sample because contributions are only made to the sum when negative samples have higher scores, i.e. are to be placed before the positive example if ranked. The slack parameter can be learned, but our experiments indicate that the model is not sensitive to this value and we use $\epsilon = 1$.
The loss function is then defined as:
\begin{align}
    L &= \log(1 + r_{i,j})
\end{align}
The logarithm is applied because ranking a positive sample with $r=1, 000$ is almost as bad as $r=100,000$ from a user perspective. In pairwise ranking methods such as BPR, where only one negative example is sampled per positive example, it is quite likely that the positive example already has a higher rank than the negative example and thus there is nothing for the model to learn. In WMRB, where multiple negative samples are used per positive example, it is much more likely that an incorrectly ranked negative example has been sampled and therefore the model can make useful parameter updates. It has been demonstrated that WMRB leads to faster convergence than pairwise loss functions and improved performance on a set of benchmark recommendations tasks \cite{Liu2017WMRB:Approach}.

\subsection{Metrics}

Each model of hyperbolic space has a distance metric that could be used as the basis for a hyperbolic recommender system. We use the hyperboloid model as it is the best suited to stochastic gradient descent-based optimisation because it is unbounded and has closed form RSGD updates. These factors have been shown in previous work to lead to significantly more efficient optimisers~\cite{Wilson2018, Nickel2018}.
The hyperboloid distance is given by
\begin{align}
d(\vec{u}, \vec{v}) &= \arccosh\left(-\left<\vec{u},\vec{v}\right>_{\mathbb{H}}\right).
\end{align}
As $\arccosh(x)$ is not defined for $x<1$ and $\left<u,v\right>_{\mathbb{H}} > -1$ can occur due to numerical instability, care must be taken within the optimiser to either catch these cases, or use suitably high precision numbers (see~\cite{DeSa2018}). In addition, the derivative of the distance 
\begin{align}
    \frac{\partial d}{\partial \vec{v}} &=
    \frac{g\vec{u}}{\sqrt{\left<\vec{u},\vec{v}\right>_{\mathbb{H}}^2-1}},
\end{align}
has the property that $\nicefrac{\partial d}{\partial \vec{v}} \to \infty$ as $\vec{u} \to \vec{v}$ because $\left<\vec{u},\vec{v}\right>_{\mathbb{H}} \to -1$. In the asymmetric framework, this is guaranteed to happen to all users that have interacted with only a single item. To protect against infinities, it is possible to use a small margin $\epsilon = 1\times10^{-6}$ leading to a distance function of
\begin{align}
d(\vec{u}, \vec{v}) &= \arccosh\left(-\left(\left<\vec{u},\vec{v}\right>_{\mathbb{H}}+\epsilon\right)\right).
\end{align}

As the hyperboloid distance is a monotone function of the Minkowski inner product and our objective is to rank points, the two are interchangeable. We generally favour the inner product as the gradient does not contain a singularity at $d=0$ and it is faster to compute.


\subsection{Optimiser}

The optimiser uses RSGD to perform gradient descent updates on the hyperboloid. There are three steps: (1) the inverse Minkowski metric $g$ is applied to Euclidean gradients of the loss function $L$ to give Minkowski gradients $\vec{h}_m$ (2) $\vec{h_m}$ are projected onto the tangent space $T_{\vec{x}}\mathbb{H}^n$ to give tangent gradients $\vec{h}_t$ (3) points on the manifold $\vec{x}$ are updated by mapping from the tangent space to the manifold with learning rate $\lambda$ through the exponential map $\mathrm{Exp}_{\vec{x}_t}$:
\begin{align}
    \vec{h}_m &= g^{-1}\nabla_\vec{x} L \\
    \vec{h}_t &= \Pi_{\vec{x}}(\vec{h}_m) \\
    \vec{x}_{t+1} &= \mathrm{Exp}_{\vec{x}_t}(-\lambda \vec{h}_t).
\end{align}

Additionally, points must be initialised on the hyperboloid. Previous work has either mapped a cube of Euclidean points in $\mathbb{R}^n$ to the hyperboloid by fixing the first coordinate~\cite{Nickel2018} or initialised within a small ball around the origin of the Poincar\'e ball model and then projected onto the hyperboloid~\cite{Wilson2018}. We find that optimisation convergence can be accelerated by randomly assigning points within the Poincar\'e ball (prior to projection to the hyperboloid), but sampling the radius $r \sim \nicefrac{1}{\log{n_i}}$ where $n_i$ is the frequency of occurance of item $i$ in the training data.
Finally, we also apply some gradient norm clipping to the tangent vectors $\vec{h}_t$.

\subsection{Item Aggregation}

We are inspired by~\cite{Steck2015a}, where model complexity is reduced by eliminating the need to learn an embedding layer for users. Instead, vectors for users $v_u$ are computed as an intermediate representation by aggregating the vectors $v_i$ of the set $I_u$ of items they have interacted with. In Euclidean space, this can be done simply by taking the mean: $v_u = \sum_{i \in I_u} \alpha_i v_i$ where $\alpha_i$ are a set of weights and in the simplest case $\alpha_i = \nicefrac{1}{|I_u|}$.

As hyperbolic space is not a vector space, an alternative procedure is required. A choice suitable for all Riemannian manifolds is the Fr\'{e}chet mean \cite{Frechet1948,Arnaudon2013}, which finds the center-of-mass, $\vec{p}$, of a cluster of points, $\vec{x}_i$, using the Riemannian distance, $d$. 
\begin{equation}
    \argmin_{\vec{p}\in\mathcal{M}}\sum_i^N d^2\left(\vec{p}, \vec{x}_i\right).
\end{equation}
The Fr\'{e}chet mean is not directly calculable, but must be found through an optimisation procedure. Despite fast stochastic algorithms, this must be recalculated for every training step and would dominate the runtime. 

To avoid this computational burden, we exploit the relationship between the hyperboloid model and the Minkowski spacetime of Einstein's Special Theory of Relativity. Given the Lorentz group of isometry-preserving group actions, we can aggregate the user-item-history by directly calculating the relativistic center-of-mass (treating all items as having unit mass). This center-of-mass is analogous to the ``Einstein midpoint'' \cite{Ungar2009}, which is most efficiently calculated in the Klein model, following projection from the hyperboloid. The midpoint is given by

\begin{equation}
  \vec{p} = \frac{\sum_i \gamma_{\vec{x}_{i}}\vec{x}_{i}}{\sum_i \gamma_{\vec{x}_i}},
  \end{equation}
  where
  \begin{equation}
  \gamma_{x_i} = \frac{1}{1-\vec{\lVert x_i \rVert}^2}.
\end{equation}

\begin{figure}[ht]
\centering
\includegraphics[width=0.44\textwidth]{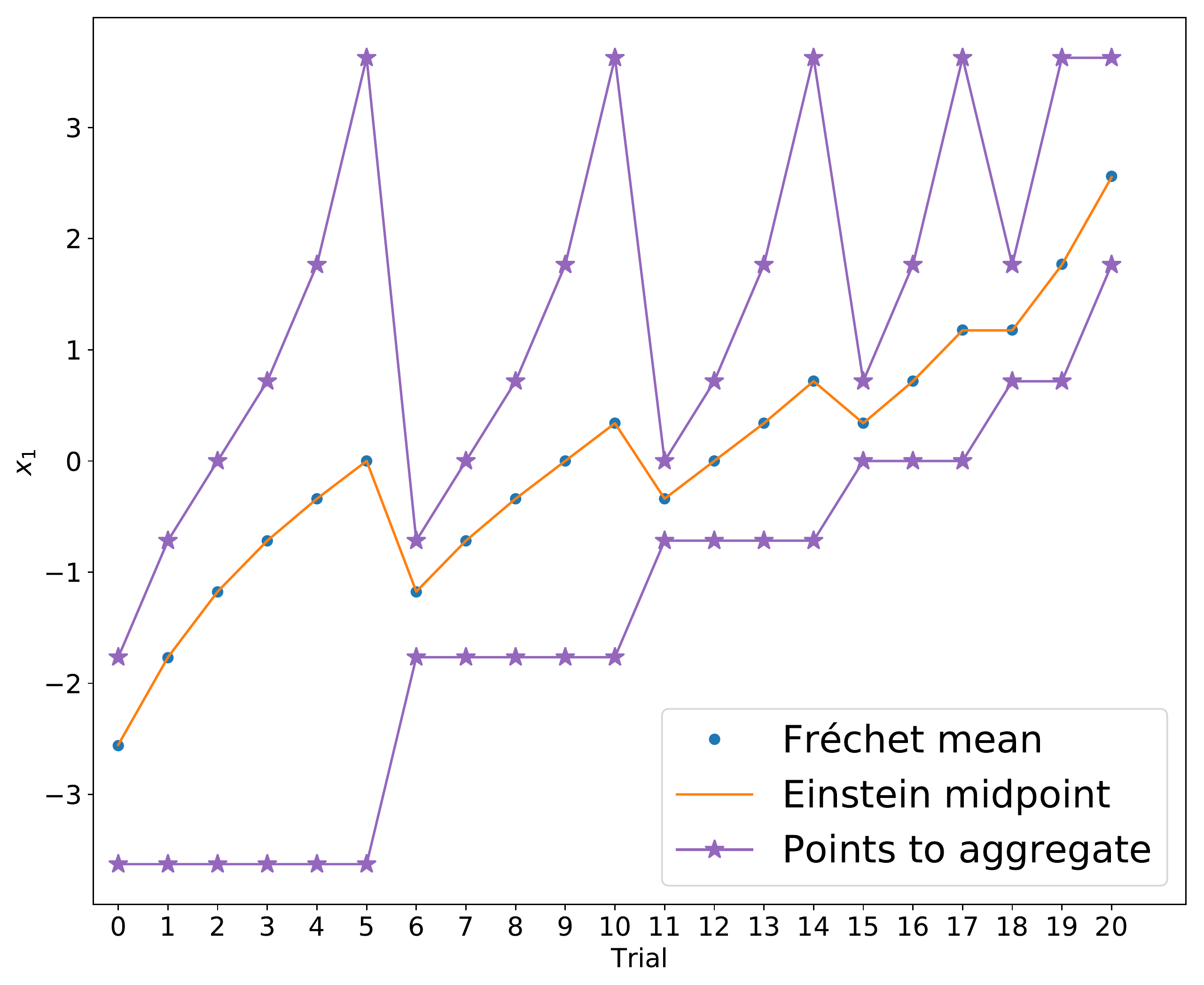}

\caption{Comparison of the aggregation of trial points (purple stars) using Einstein midpoints (orange line) to Fr\'echet means (blue points).}
\label{fig:frechet_vs_einstein}
\end{figure}

Figure~\ref{fig:frechet_vs_einstein} shows a comparison of the Einstein midpoint to the Fr\'{e}chet mean for a scan over $\alpha: -2 \leq \alpha \leq 2$, where $\vec{x} = (\sinh(\alpha), \cosh(\alpha))$. For each initial point all greater values of $\alpha$ were compared. The Fr\'{e}chet mean optimisation used gradient descent, for ten iterations. Agreement better than 0.3\% is observed for all the aggregation points tested, with the precision limited by the number of gradient descent steps performed. Due to the close agreement and superior runtime complexity, 
the Einstein midpoint is used for item aggregation.

\section{Evaluation}

We report results from experiments on simulations, eight Amazon review datasets, the MovieLens 20M~\cite{harper2016movielens}, and finally a large ASOS proprietary dataset. Each experiment represents a milestone towards the development of full-scale hyperbolic recommender systems.
Experiments report the standard recommender system metrics Hit Rate at 10 and Net Discount Cummulative Gain at 10, which we denote as HR@10 and NDCG@10 respectively.

\subsection{Simulations}

To demonstrate the viability of hyperbolic recommender systems, we present three small scale simulations. The simulations are generated using a symmetric, hyperboloid recommender with explicit user representations and the BPR loss. The embeddings are then projected onto the Poincar\'e disk to product the figures.

\begin{figure*}
    \centering
    \includegraphics[width=.25\textwidth]{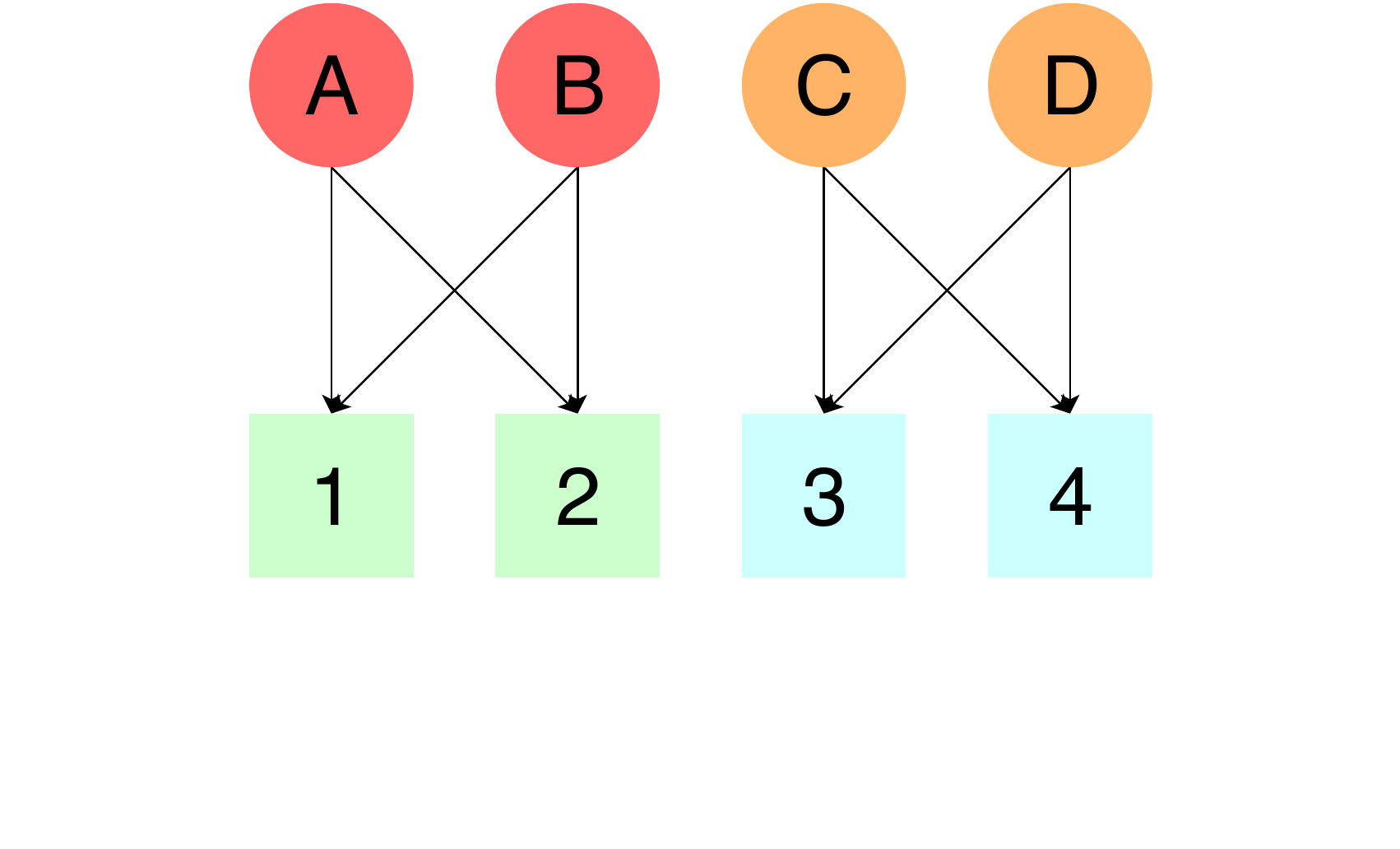}\hfill
    \includegraphics[width=.25\textwidth]{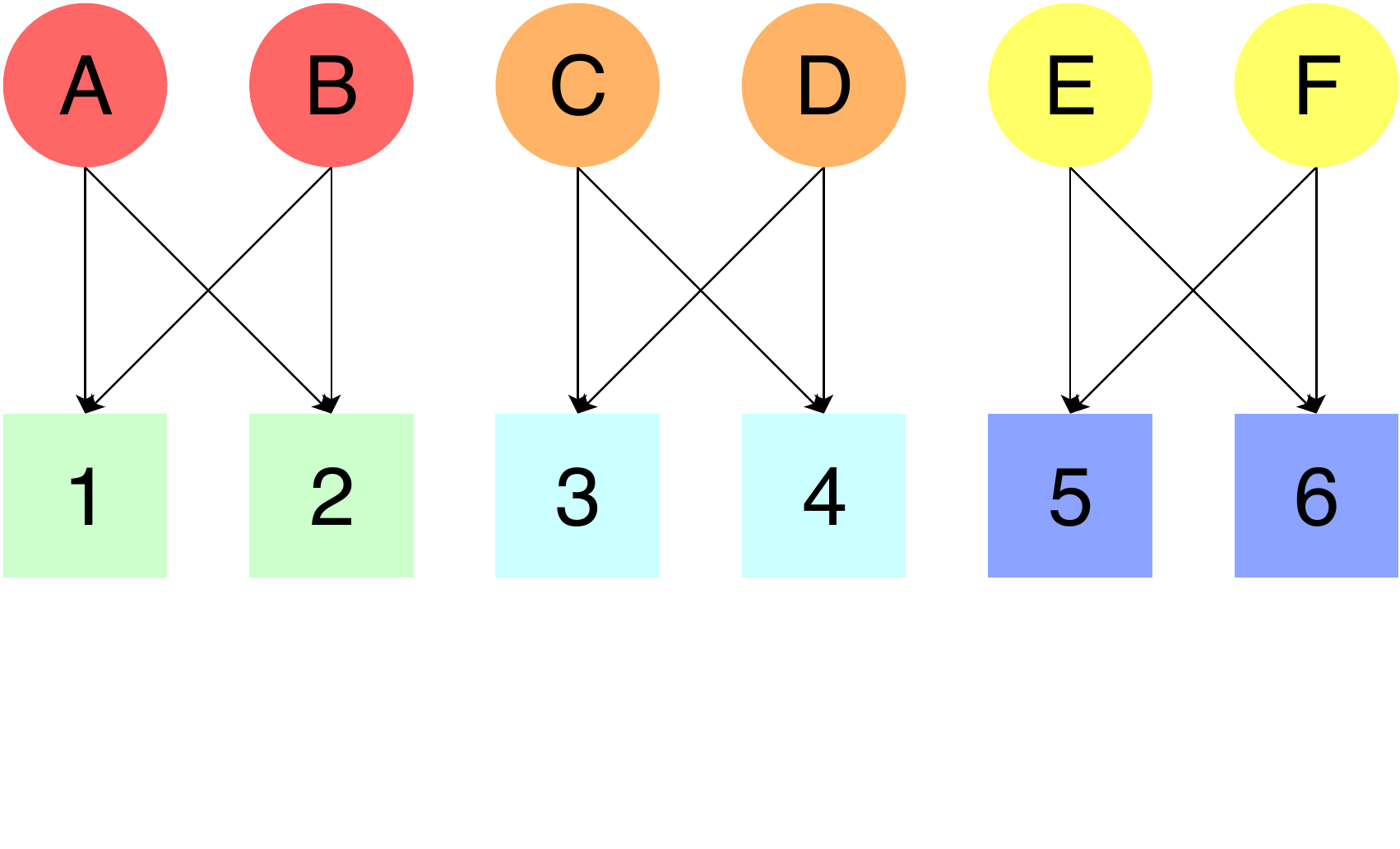}\hfill
    \includegraphics[width=.25\textwidth]{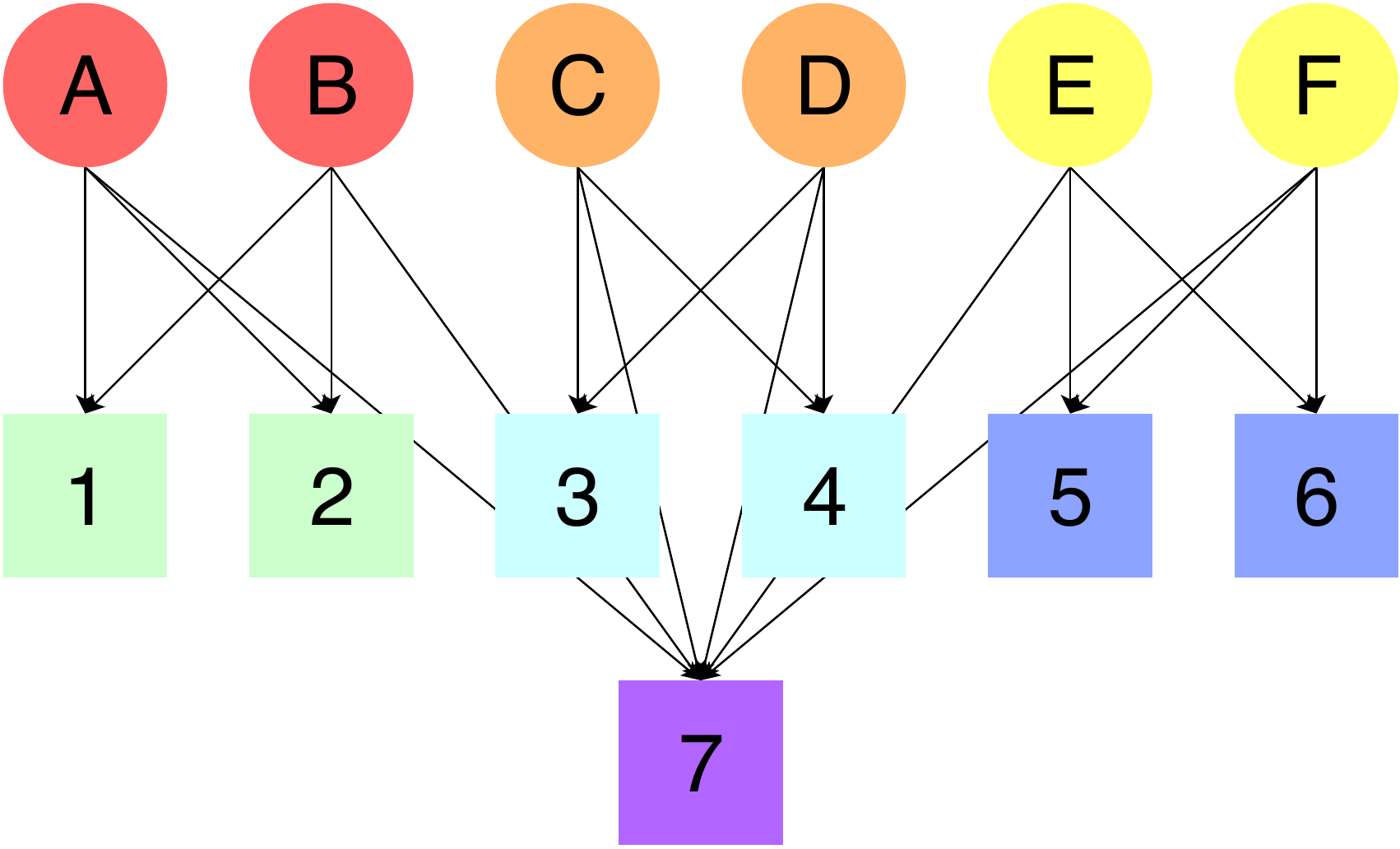}

    \includegraphics[width=.25\textwidth]{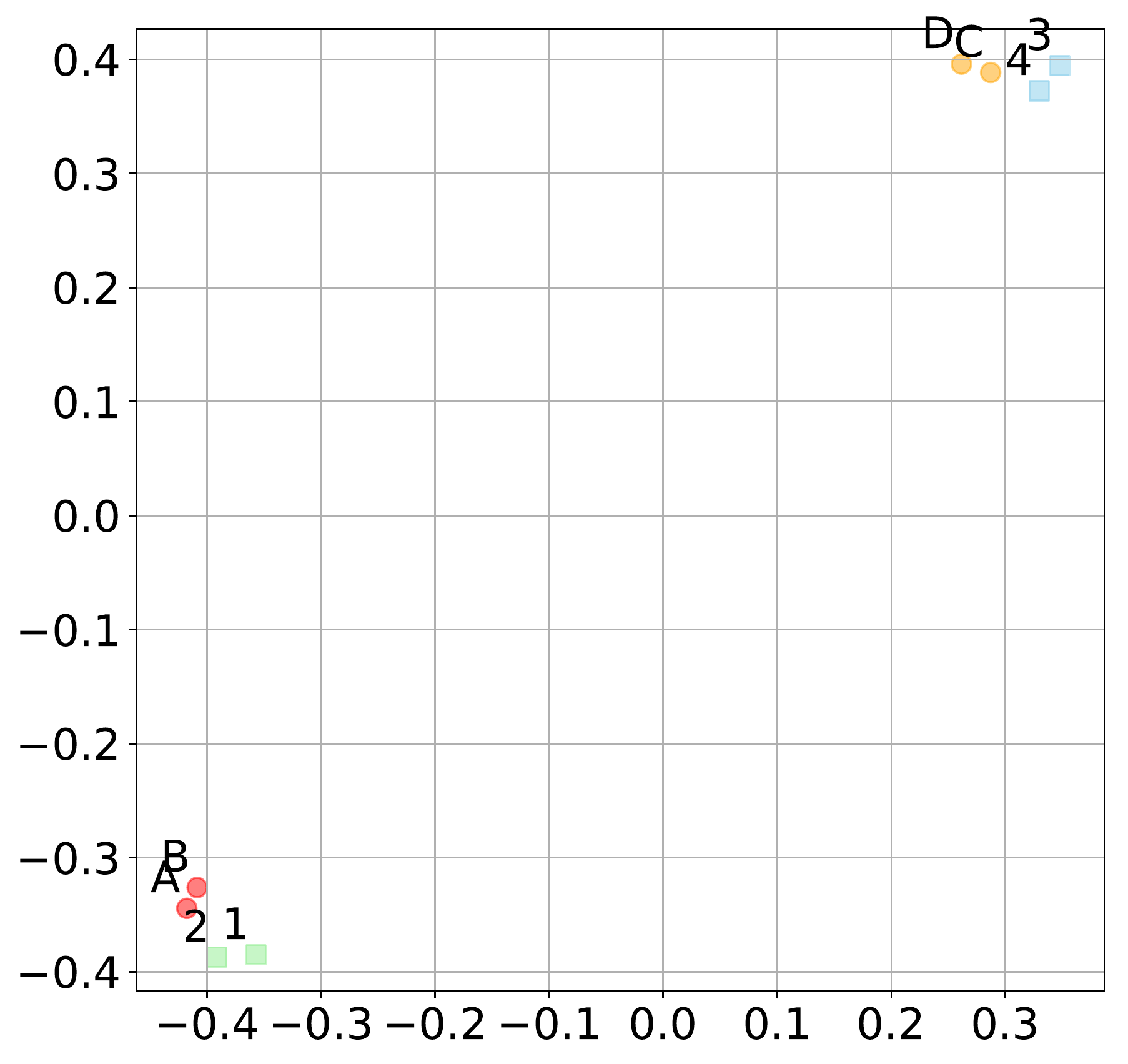}\hfill
    \includegraphics[width=.25\textwidth]{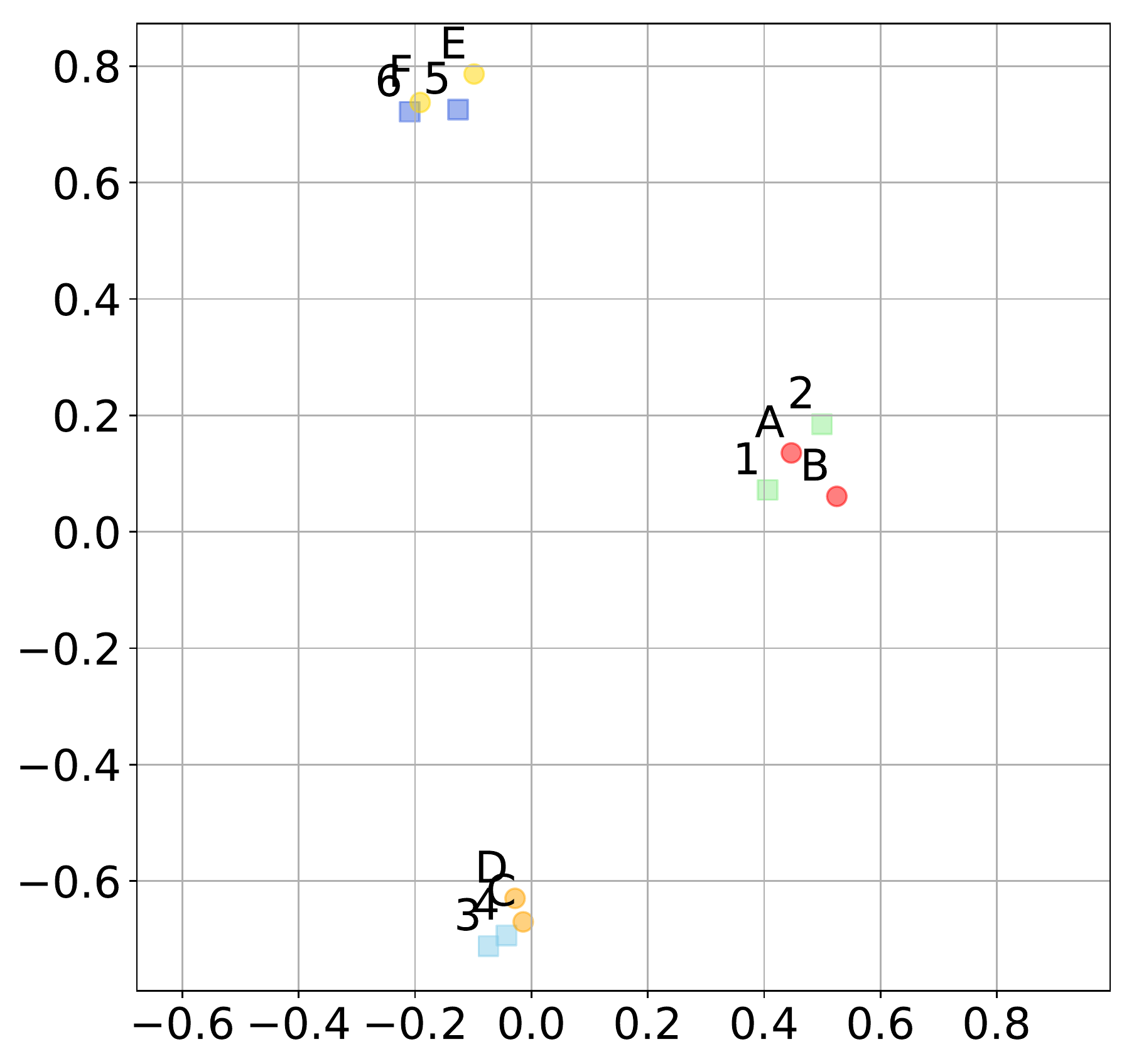}\hfill
    \includegraphics[width=.25\textwidth]{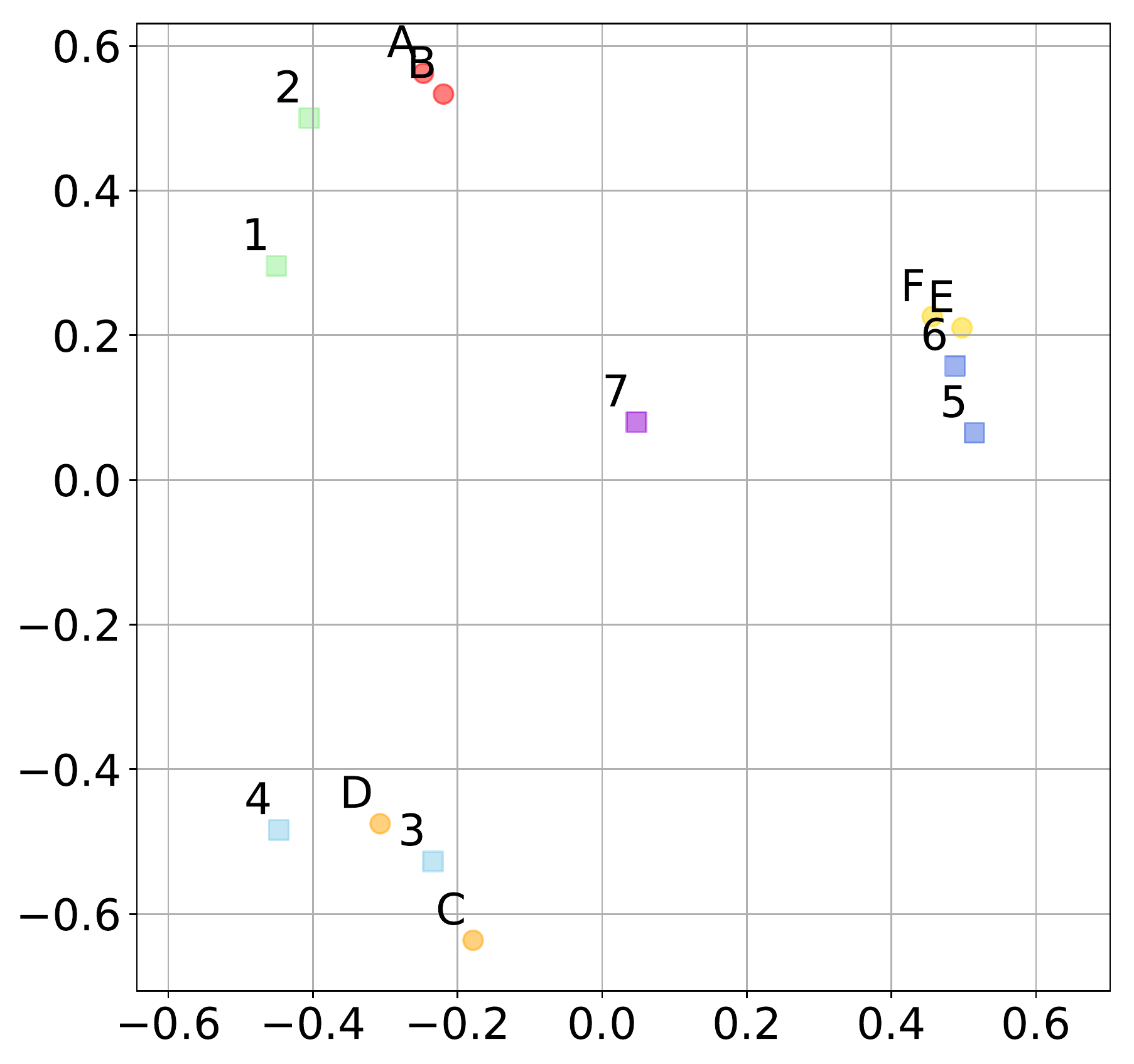}\hfill
    \caption{Simulation experiments for user-item graphs. Users are represented as circles with alphabetical IDs, while items are represented as squares with numeric IDs. Top row: Three simple user-item graphs, with edges between a user and an item representing a purchase. Bottom row: The corresponding embeddings of users and items on a 2D hyperboloid, projected onto the Poincar\'e disk for visualisation}
    \label{fig:simulation}
\end{figure*}

The first simulation (Figure~\ref{fig:simulation}, left column) consists of four users and four items clustered into two disjoint bipartite graphs
An effective recommender system embeds users close to items they have purchased and distant from items they have not purchased. Therefore, we would expect the final embeddings to consist of two distinct groups, with users A and B and items 1 and 2 all embedded very close to one another, and users C and D and items 3 and 4 also embedded close to one another, but a large distance away from the first group. As can be seen, this is exactly what is learned by the symmetric hyperboloid recommender system. 

In the second simulation (Figure~\ref{fig:simulation}, middle column), a third disjoint user-item graph is added to the system. Again, users and items within each group share very similar embeddings, with high inter-group separations. 

In the third simulation (Figure~\ref{fig:simulation}, right columns), An additional item, labelled 7, that has been purchased by all six users is added. Consequently, an effective recommender system will learn a set of embeddings such that item 7 is close to all six users, while still maintaining a distance between users in each group and items that were bought exclusively by members of one of the other groups. As can be seen, the resulting set of embeddings learned by the symmetric hyperboloid recommender system is very similar to those produced in the second simulation, however, item 7 is embedded near the origin. This is consistent with previous work embedding tree-like graphs~\cite{Nickel2017, Chamberlain2017e} as item 7 is effectively higher up the product hierarchy than items 1-6. This simple system highlights the strengths of using hyperbolic geometry for recommendations. All users are equally close to item 7, however due to the geodesic structure of the Poincar\'e ball, they are still a large distance away from the users and items in the other groups. 



\subsection{Amazon Review Datasets}

\begin{figure*}[ht]
\centering
\begin{tabular}{c@{}c@{}c@{}c@{}}
\includegraphics[width=0.24\textwidth]{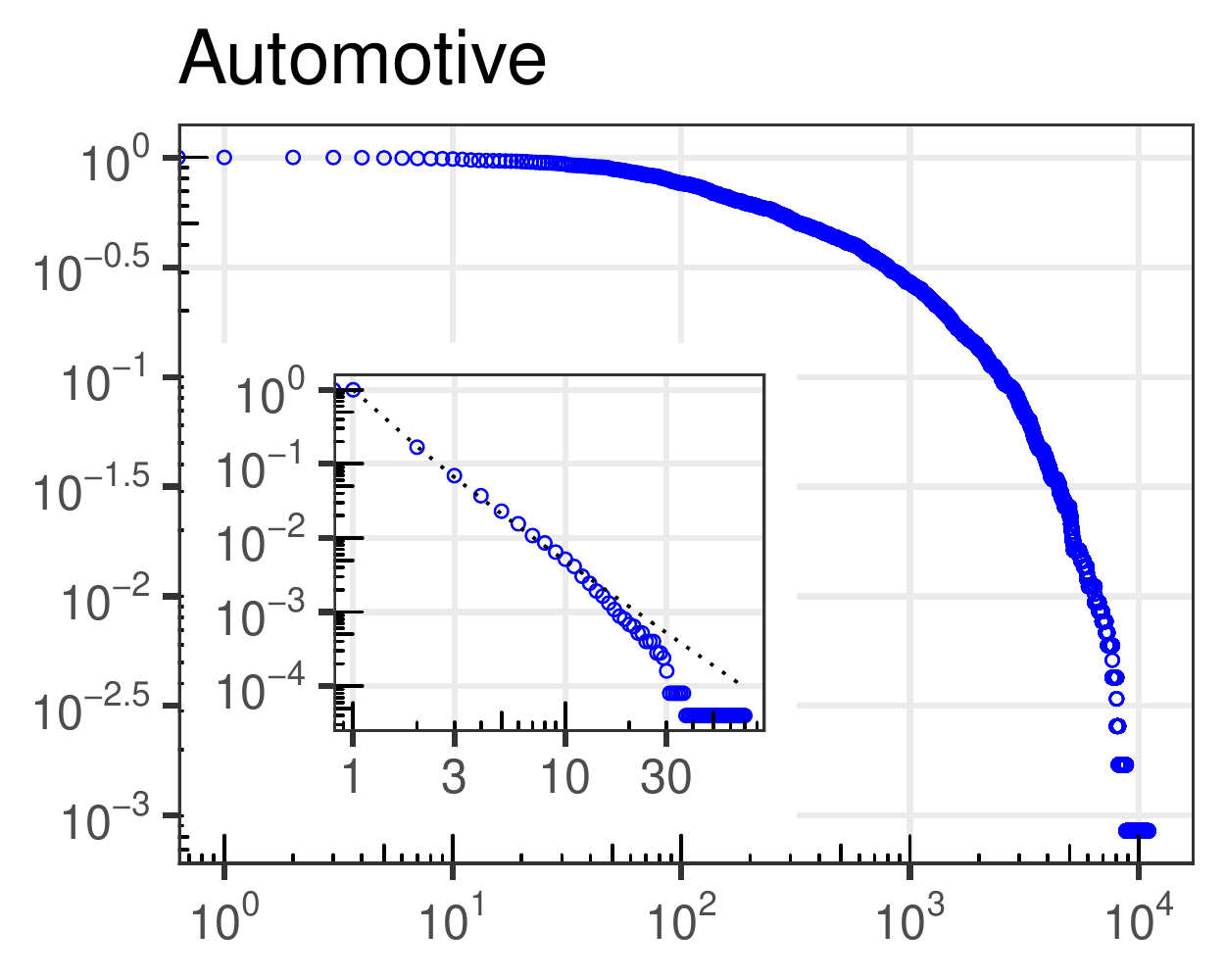} &\includegraphics[width=0.24\textwidth]{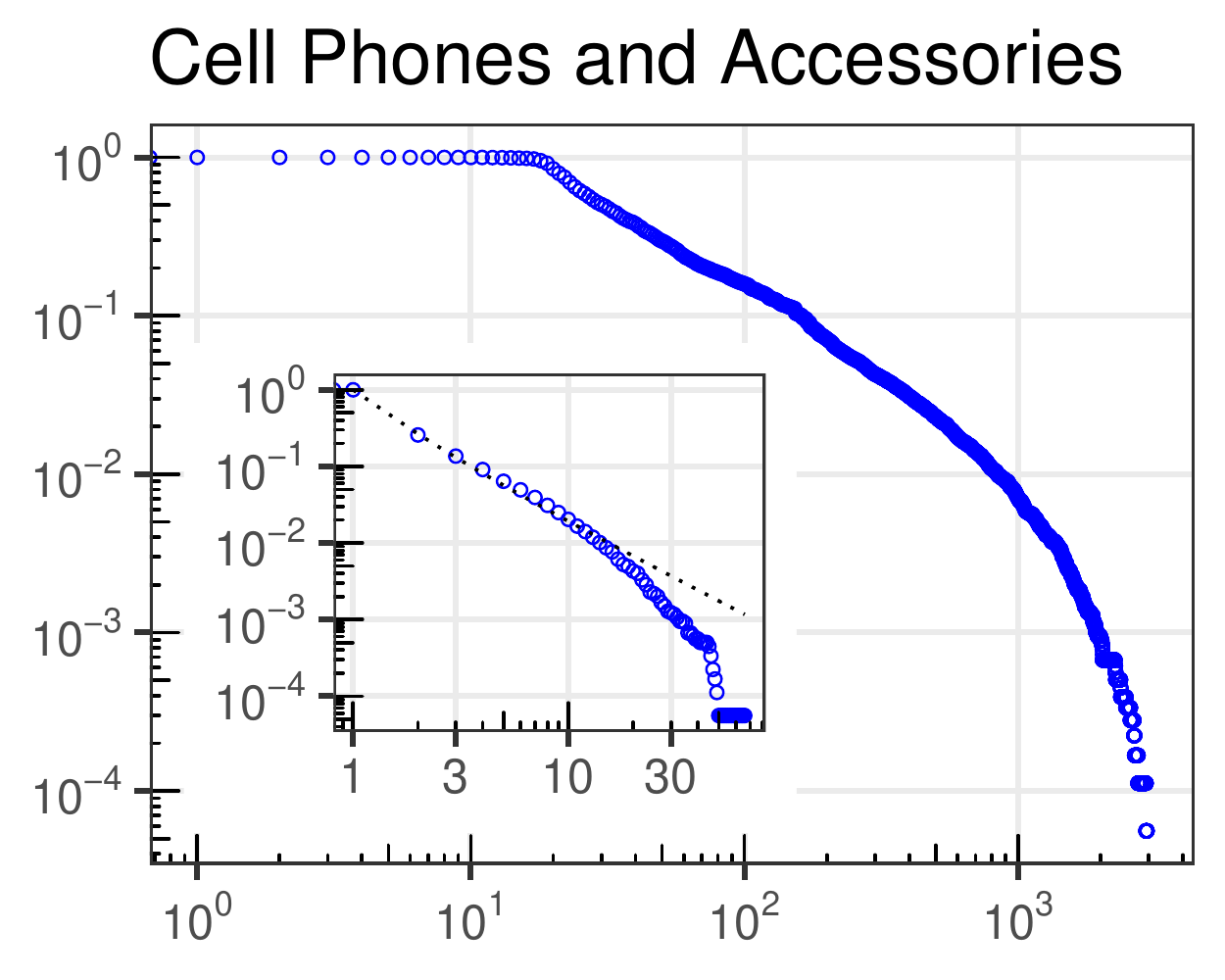}
&\includegraphics[width=0.24\textwidth]{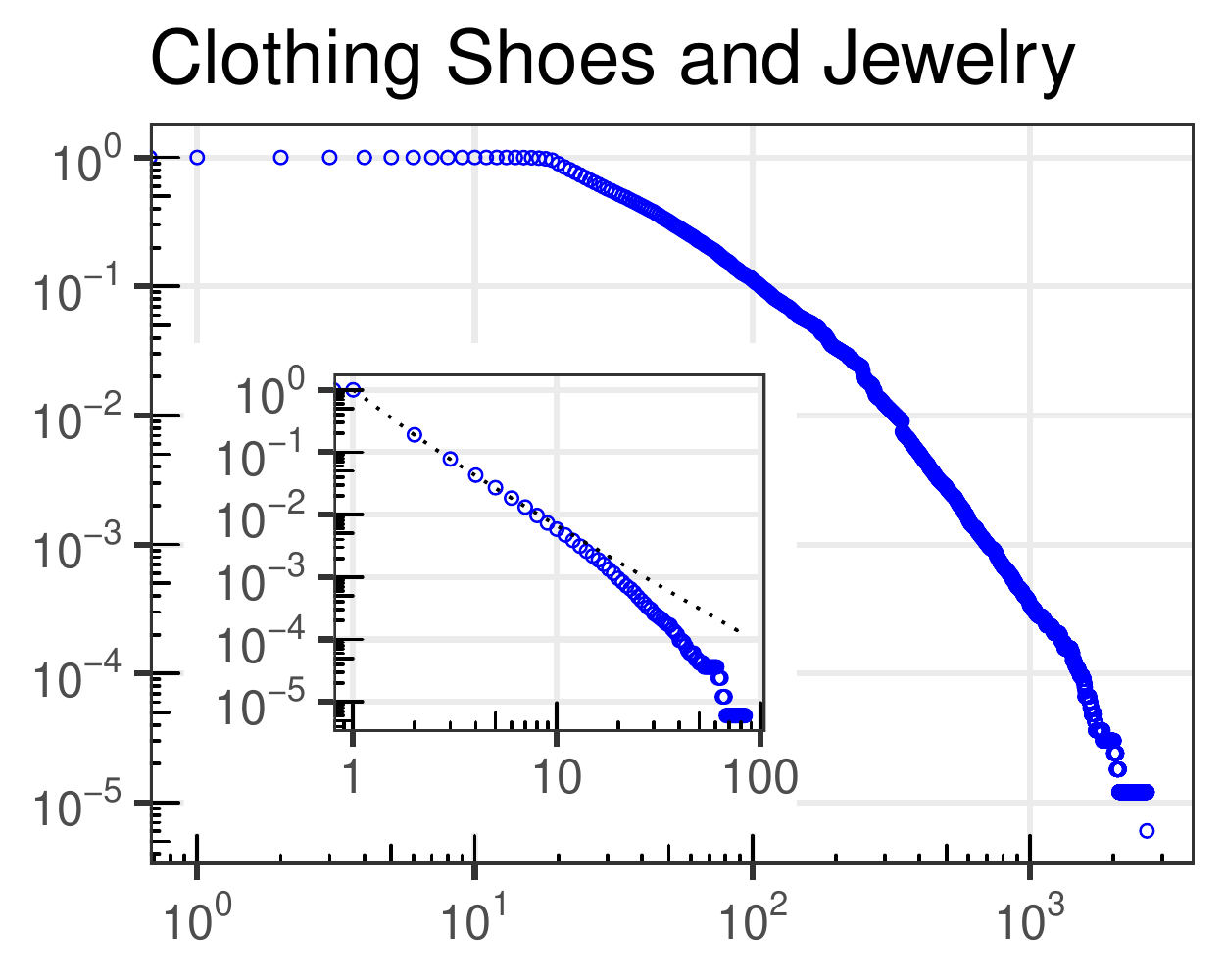} &\includegraphics[width=0.24\textwidth]{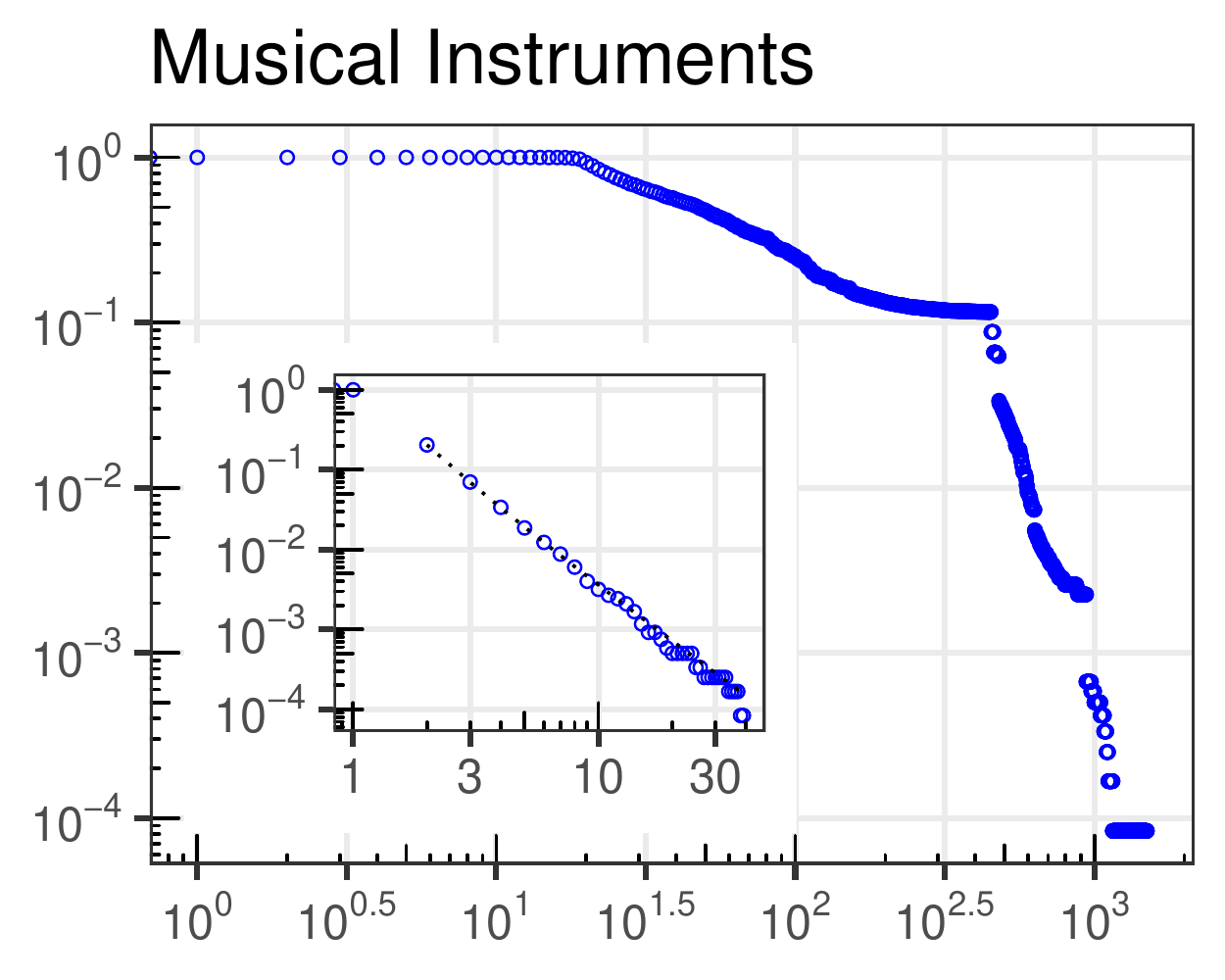}\\
\includegraphics[width=0.24\textwidth]{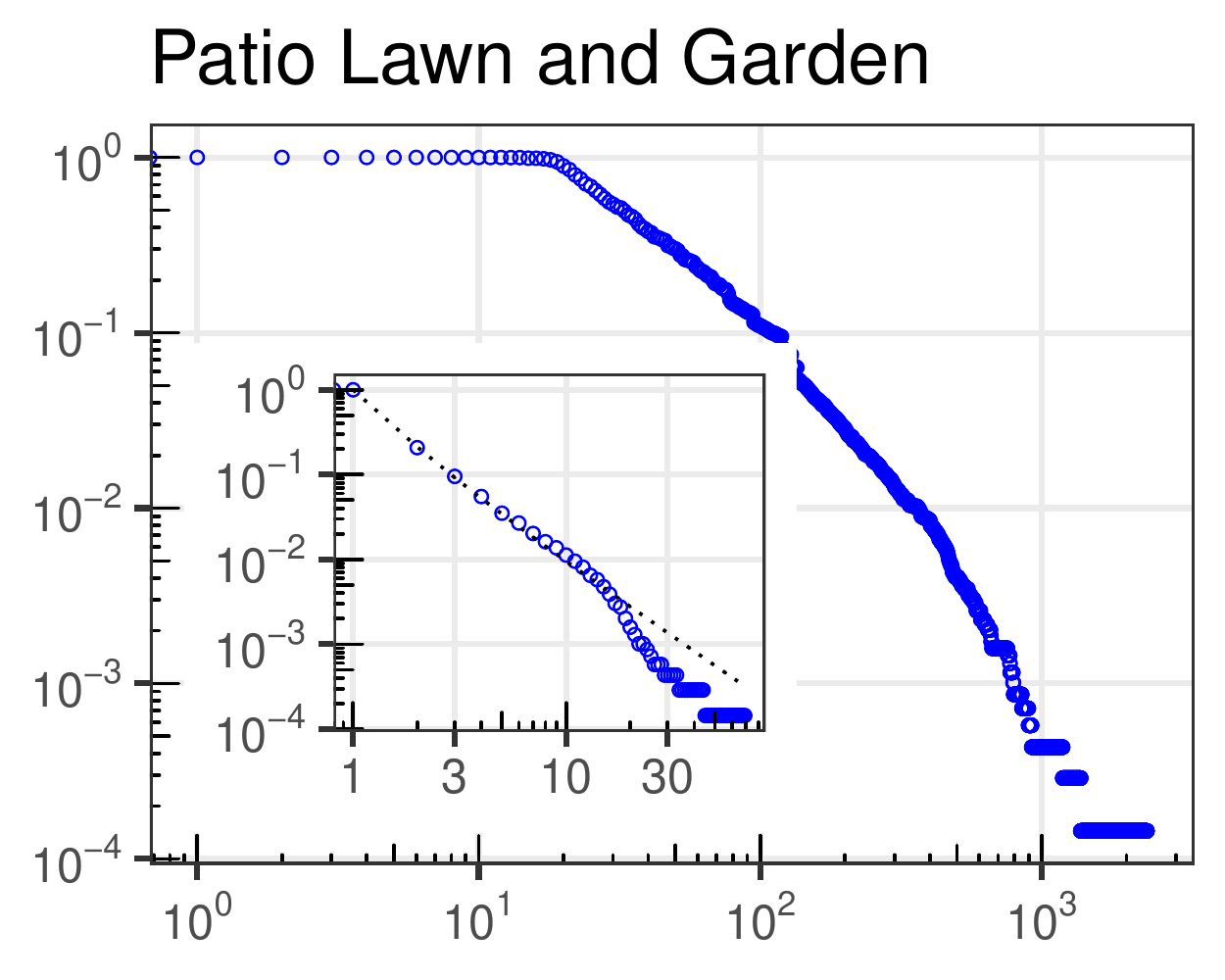} &\includegraphics[width=0.24\textwidth]{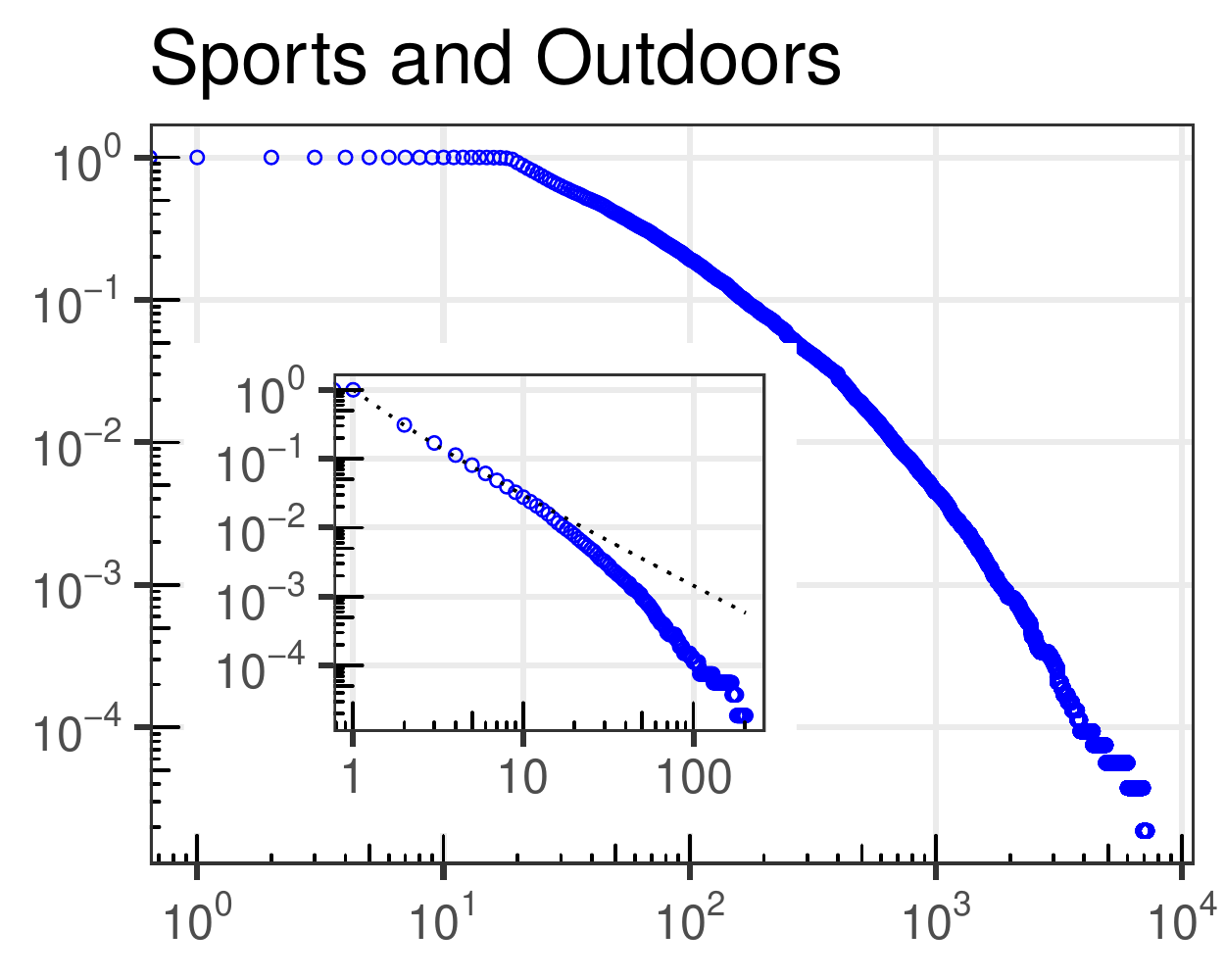}
&\includegraphics[width=0.24\textwidth]{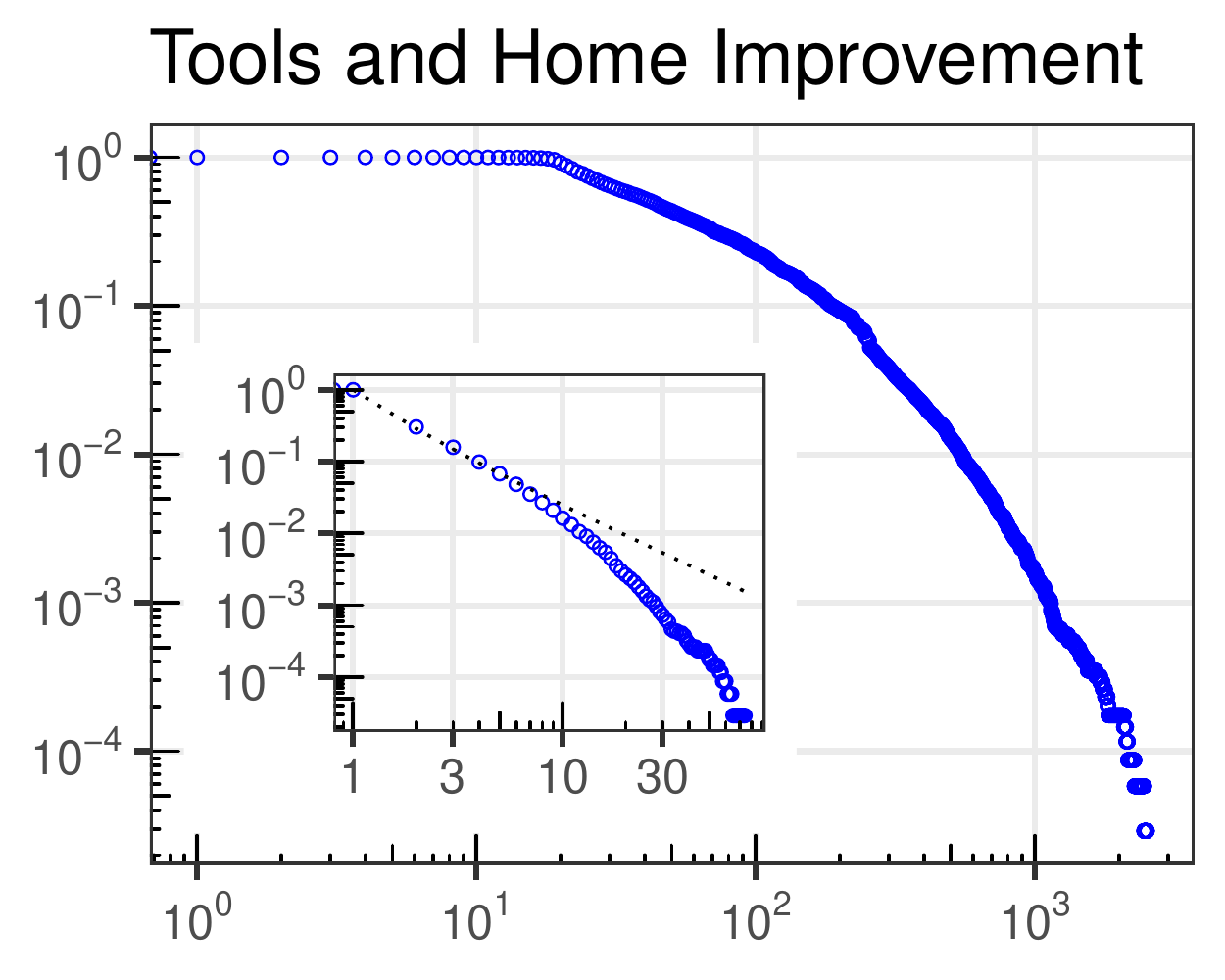} &\includegraphics[width=0.24\textwidth]{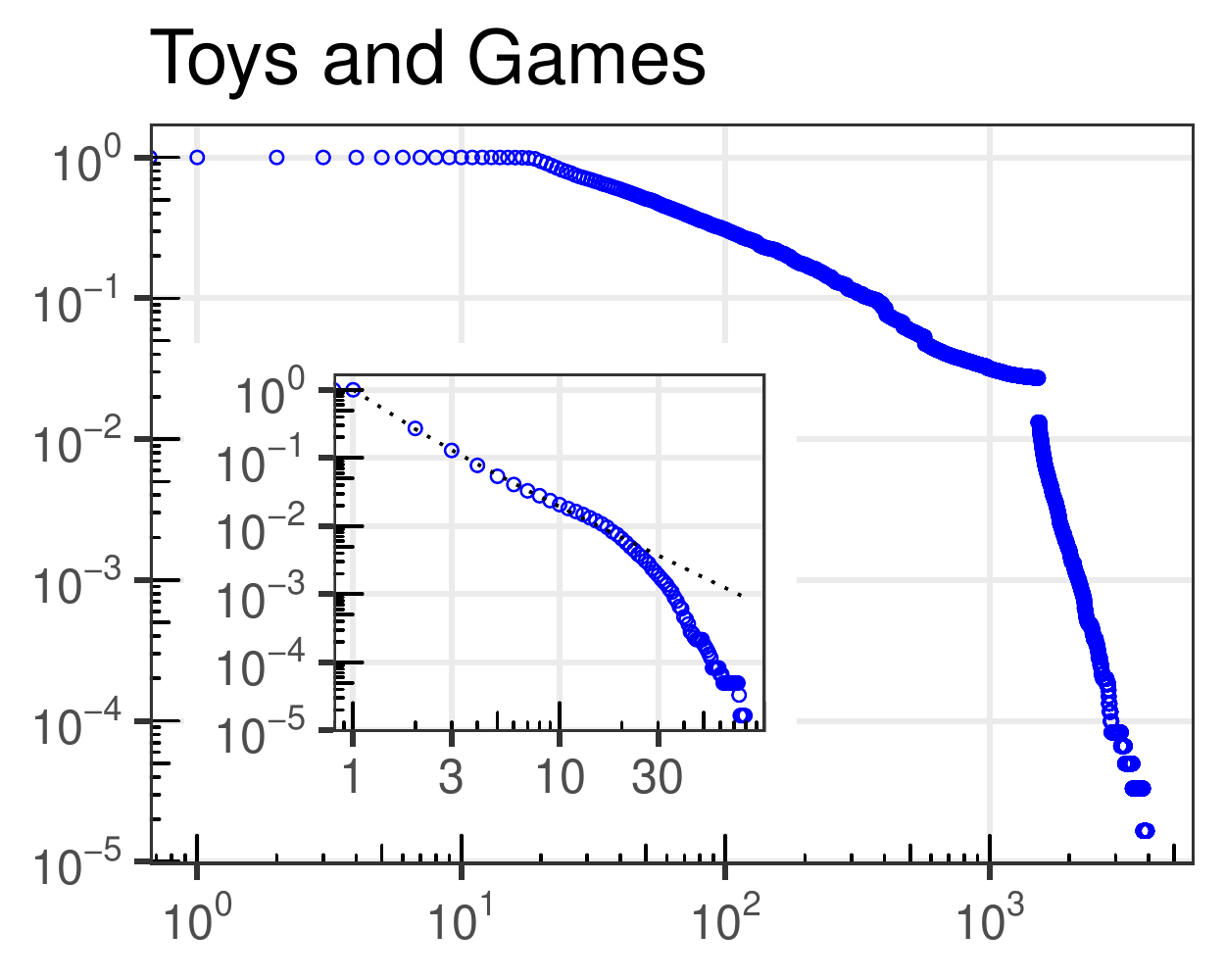}
\end{tabular}
\vskip -0.1in
\caption{Log-log degree distributions of interaction graphs from the Amazon data sets. $x$-axis: the degree value, $y$-axis: the empirical frequency of nodes with degree higher than $x$. Main plot: cumulative distribution of the weighted degree in the item-to-item projection, where two items are connected if the same customer interacts with them. Inset: cumulative degree distribution of item nodes in the customer-to-item bipartite graph. Dotted lines in the insets give the maximum-likelihood estimate of a power-law fit.}
\label{fig:dd-amazon}
\end{figure*}

Having demonstrated the viability of hyperbolic recommender systems for small simulations, we apply the same symmetric, hyperboloid BPR based model to the Amazon Review datasets and show that it outperforms the equivalent Euclidean model. 
We choose the Amazon datasets as our analysis of the underlying networks, presented in Table~\ref{tab:stats-nets} and Figure~\ref{fig:dd-amazon}, shows that these datasets are examples of complex networks.


Euclidean and hyperbolic methods are evaluated by training on all interactions from users with more than 20 interactions. The final performance is assessed on a held out test set composed of the most recent interaction each user has had with an item using HR@10 with 100 negative samples.



To ensure our benchmark emphasises the difference in the underlying geometry in the task of user-item recommendation, hyperparameter tuning for both Euclidean and hyperbolic models follows an identical procedure. The dimensionality of the embedding is at 50 and we search for optimal learning rates and regularization parameters for each geometry over \small $\{1.0,0.8,0.5,0.1\}$, \small $\{10^{-1},10^{-2},10^{-3},10^{-4},10^{-5},10^{-6},10^{-7}\}$ \normalsize and \small $\{1.0,0.8, 0.5,0.1\}$ \normalsize
respectively. In all experiments we fix the batch size to 128 training samples, and use stochastic gradient descent.

\begin{table}[h]
\centering
  \caption{Test performance of hyperboloid and Euclidean recommenders on Amazon data. HR@10 is averaged over 9 runs, $^{(*)}$ indicate significant difference in means at 5\% level.}
  \label{tab:perf-amazon}
  \vskip -0.1in
\begin{tabular}{llll}
\toprule
\multicolumn{1}{l}{Data set} & & \multicolumn{1}{l}{Hyperboloid} & \multicolumn{1}{l}{Euclidean}
\\ 
\midrule
\multicolumn{1}{l}{Automotive}  & & $\textbf{0.59}\pm\textbf{0.01}^{(*)}$& $0.54\pm0.01$\\ 
\multicolumn{1}{l}{Cell Phones and Accessories} & & $\textbf{0.49}\pm\textbf{0.01}$ &$0.48\pm0.01$ \\ 
\multicolumn{1}{l}{Clothing Shoes and Jewelry} && $\textbf{0.59}\pm\textbf{0.00}^{(*)}$&$0.52\pm0.00$ \\ 
\multicolumn{1}{l}{Musical Instruments} &&$\textbf{0.45}\pm\textbf{0.01}$ & $0.44\pm0.01$ \\ 
\multicolumn{1}{l}{Patio Lawn and Garden} &&$\textbf{0.52}\pm\textbf{0.02}$ &$0.51\pm0.02$  \\ 
\multicolumn{1}{l}{Sports and Outdoors} && $\textbf{0.60}\pm\textbf{0.01}^{(*)}$&$0.52\pm0.00$  \\ 
\multicolumn{1}{l}{Tools and Home Improvement} && $\textbf{0.54}\pm\textbf{0.01}^{(*)}$& $0.46\pm0.00$\\ 
\multicolumn{1}{l}{Toys and Games}&& $\textbf{0.60}\pm\textbf{0.01}^{(*)}$&$0.55\pm0.01$  \\ 
\bottomrule 
\end{tabular} 
\label{tab:amazon}
\end{table} 

For each system and each dataset, the optimal learning rate and regularization value is established by repeating each experiment for $N$ runs, and assessing the average HR@10, where we have used $N=9$. Results are presented in Table~\ref{tab:amazon}. In all cases, we observe superior performance using hyperbolic geometry. 

\begin{figure}[h]
\centering
\includegraphics[width=\columnwidth,clip]{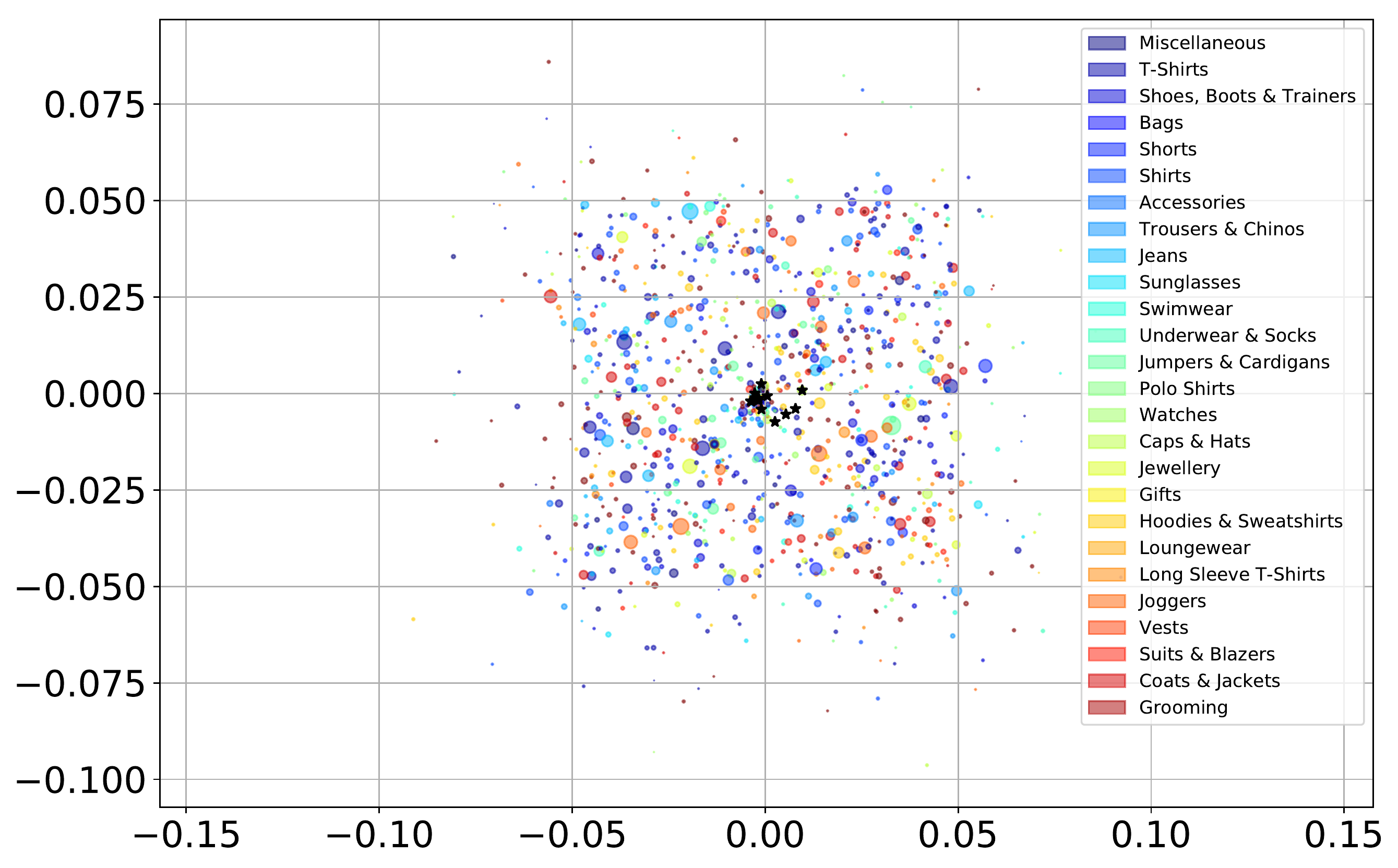}
\caption{Two dimensional hyperbolic embeddings for a sample of the ASOS dataset. Colour indicates product type while the size of points indicates item popularity. The black stars show implicit customer representations.}
\label{fig:asos_hyp_embedding}
\end{figure}

\subsection{MovieLens20M Dataset}

\begin{figure}[ht]
\centering
\includegraphics[width=0.48\textwidth]{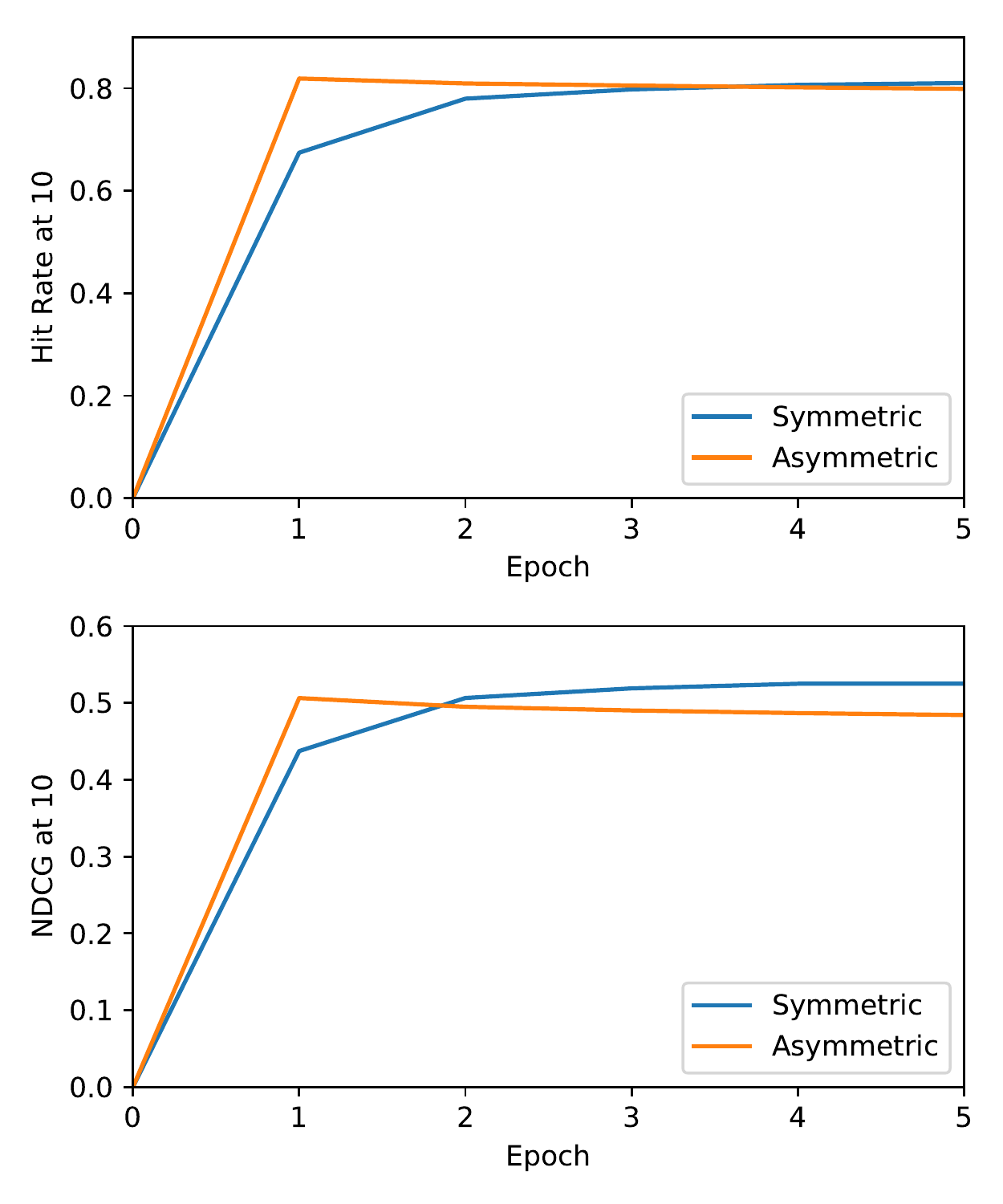}
\vspace{-0.5cm}
\caption{Comparison of the validation HR@10 and NDCG@10 with 100 negative examples, for asymmetric (orange) and symmetric (blue) hyperboloid recommender systems on the MovieLens20M dataset.}
\label{fig:asym_v_sym}
\vspace{-5mm}
\end{figure}

Given that hyperbolic recommendation systems outperform their Euclidean equivalents on datasets that have the structure of complex networks, the next milestone is to show that hyperbolic recommender systems can scale. To achieve scalability we adopt the asymmetric recommender paradigm,  where customers are not represented explicitly, but as aggregates of product representations.

We assess the performance of our asymmetric hyperboloid recommender system using the MovieLens 20M dataset~\cite{harper2016movielens}, which contains integer movie ratings. To convert it into a form consisted with co-purchasing data, we filter so that only user-item pairs in which the user has given the movie a rating of 4 or 5 are considered to be positive interactions. This results in $16,486,759$ ratings from $137,765$ users of $20,720$ (See Table \ref{tab:stats-nets} for dataset statistics).  As with the Amazon dataset, we hold out each user's most recent interaction to form a test set, and use each user's second most recent interaction as a validation set. We evaluate our results using HR@10 and NDCG@10 with 100 negative examples. 

We compare the asymmetric hyperboloid recommender system with the symmetric case using an embedding dimension of 50, a learning rate of 0.01 and a batch size of 1024 with stochastic gradient descent. The performance of the asymmetric system is roughly equivalent to the symmetric system, but the asymmetric system is able to learn in half the time using five times less parameters (Figure~\ref{fig:asym_v_sym}). Fast convergence is important in production recommender systems, where large datasets containing millions of users are retrained daily. 

\subsection{Proprietary Dataset}

Finally, we assess the performance of the hyperboloid recommender system on an ASOS proprietary dataset, which consists of 28m interactions between $O(10^6)$\footnote{the exact number can not be disclosed for commercial reasons}  users with $132,399$ items over a period of one year. Embeddings for a sample of this dataset in 2D hyperbolic space and projected into the Poincar\'e disc is shown in Figure~\ref{fig:asos_hyp_embedding}. In the figure points are coloured by product type and scaled by item popularity with black stars showing the implicit customer representations.

In these experiments, the test set consisted of the last week of interactions, with the training and validation sets formed from the previous 51 weeks of data. The validation set consisted of $500,000$ interactions drawn uniformly at random in time, with the remainder forming the training set.

In all configurations, the runtime of the symmetric system was four times greater than the asymmetric system for a fixed number of epochs.
%
With 50 embedding dimensions, learning rate of 0.05, batch size of 512 and training for a single epoch, we observed a test set HR@10 = 0.589 and NDCG@10 = 0.324, significantly better than random and demonstrating that the system can learn on large commercial datasets. However, this performance was worse than the equivalent Euclidean asymmetric recommender, which gave HR@10 = 0.639 and NDGC@10 = 0.393, when trained with the same hyperparameters. 

Although the performance of the hyperboloid recommender did not surpass that of the Euclidean-based system, we believe these results are extremely promising.  
The performance of the hyperboloid recommender system could be significantly improved by applying adaptive learning rates, particularly through development of adaptive optimisation techniques that function on the hyperboloid. Improvements to the initialisation scheme used should also be investigated. 
\section{Conclusion}

We have presented a novel hyperbolic recommendation system based on the hyperboloid model of hyperbolic geometry. Our approach was inspired by the intimate connections between hyperbolic geometry, complex networks and recommendation systems. We have shown that it consistently and significantly outperforms the equivalent Euclidean model using a popular public benchmark. We have also shown that by using the Einstein midpoints, it is possible to develop asymmetric hyperbolic recommender systems, which can scale to millions of users, achieving the same performance as symmetric systems, but with far fewer parameters and greatly reduced training times. We believe that future work to develop adaptive optimisers in hyperbolic space will lead to state-of-the-art production-grade hyperbolic recommender systems.

\FloatBarrier
\bibliographystyle{ACM-Reference-Format}
\newcommand{\showDOI}[1]{\unskip}
\newcommand{\showURL}[1]{\unskip}
\renewcommand{\showeprint}[8]{\unskip}
 \bibliography{main.bib}
\pagebreak
\section{Reproducibility Guidance}

This section contains detailed instructions to aid in the reproduction of our experimental results. All code and data used in the experiments are available on request.

\subsection{Evaluation Metrics}

We evaluate our results using the Hit Rate (HR) at 10 and Normalised Discount Cummulative Gain (NDCG) at 10. to calculate hit rate, each positive example in the held out set is ranked along with 100 uniformly sampled negative examples that the user has not interacted with, the proportion of cases a positive example is ranked in the top 10 closest to the user ("hits") yields the performance of a system. NDCG@10 sums the relevance of the first 10 items discounted by the log of their position and normalised by the NDCG@10 of the ideal recommender.

\subsection{Simulated Experiments}

The simulations use a 2-dimensional hyperboloid, the BPR loss, a learning rate of 1, a decay rate of 0.02 and an initialisation width of 0.01.

\subsection{Amazon Review Datasets Experiments}
 All experiments were conducted using Python3 and Torch-1.0 on Ubuntu 16.04, with a Tesla 2xK80 - 16Gb Ram.

Optimal parameters for the hyperboloid model were found to be $10^{-2}$ for the learning rate, and $1.0$ for the regularisation parameter. It was found to vary in the case of the Euclidean model, with respectively $(0.1, 0.5)$ on automotive, $(0.1, 0.5)$ on cellphones, $(0.1, 1.0)$ on patio, $(0.1, 1.0)$ on clothing, $(0.1, 1.0)$ on musical, $(0.1, 0.8)$ on toys, $(0.1, 0.8)$ on tools, $(0.01, 1.0)$ on sport. The mini-batch size used was 128, with the models trained for 10 epochs. Only plain updates were considered, where the learning rate was held constant at each epoch. The test set is composed of every last positive interaction a user has had (with a rating score $>1$). Positive interactions not seen during training were removed from the test set to ensure the performance of a model only reflects interactions that were fully optimised.

\subsection{MovieLens20M Dataset Experiments}

The analysis of the asymmetric and symmetric datasets on the MovieLens20M dataset were conducted with the following parameters: Gradients were clipped in the tangent space to norm 1, learning rates were 0.1 using SGD, embedding dimension was 50 and the loss was WRMB with 100 negative samples and regularisation of 0.01. Embeddings were initialised uniformly at random into a hypercube of width 0.001 and then projected onto the hyperboloid if appropriate.

\subsection{Derivatives of the Loss Function}

Here we cover the case for the WMRB loss using the hyperboloid distance. The gradients using the inner product are the same with the $\arccosh$ derivative removed and largely similar for the BPR loss function.

The wmrb ranking loss is given by
\begin{align}
    \mathrm{rank}(i,j) &\approx r_{i,j} = \sum_{k \in N} \lvert \epsilon + d(\vec{u}_i,\vec{v}_j) -d(\vec{u}_i,\vec{v}_k) \rvert_+ 
    \end{align}
where $d(\vec{u}_i,\vec{v}_{j,k})$ is the distance between $\vec{u}_i$ and $\vec{v}_{j,k}$, $\lvert . \rvert $ is the relu function and $\epsilon \in \mathbb{R}_+$ is a slack parameter such that terms only contribute to the loss if $d(\vec{u}_i,\vec{v}_j) + 1 >  d(\vec{u}_i,\vec{v}_k)$. 
The loss function is given by
\begin{align}
    L = \log(1 + r_{i,j})
\end{align}
\begin{align}
    \frac{\partial L}{\partial v_j} &= \frac{1}{1+r_{i,j}}\frac{\partial r_{i,j}}{\partial v_j} \\
\end{align}
We denote $\eta_k = d(\vec{u}_i,\vec{v}_j) + 1 > d(\vec{u}_i,\vec{v}_k)$ as the condition that must be satisfied for updates to occur and $K = \{\eta_k = \mathrm{True} \}$, then
\begin{align}
    \frac{\partial L}{\partial \vec{v}_j} &= \frac{|K|}{1+r_{i,j}} 
    \frac{g\vec{u}_i}{\sqrt{\left<\vec{u}_i,\vec{v}_j\right>_{\mathbb{H}}^2-1}}
\end{align}
updates for $v_k$ are similarly
\begin{align}
    \frac{\partial L}{\partial \vec{v}_k} &=
    \begin{cases}
    \frac{-1}{1+r_{i,j}}
    \frac{g\vec{u}_i}{\sqrt{\left<\vec{u}_i,\vec{v}_k\right>_{\mathbb{H}}^2-1}}, \,\, \eta_k = \mathrm{True} \\
    0, \,\, \text{otherwise}
    \end{cases}
\end{align}
However, updates of $\vec{u}_i$ are more complex
\begin{align}
    \frac{\partial L}{\partial \vec{u}_i} &=
    \begin{cases}
    \sum_{k\in K}\frac{1}{1+r_{i,j}}
    \left(\frac{g\vec{v}_k}{\sqrt{\left<\vec{u}_i,\vec{v}_k\right>_{\mathbb{H}}^2-1}} - \frac{g\vec{v}_j}{\sqrt{\left<\vec{u}_i,\vec{v}_j\right>_{\mathbb{H}}^2-1}}\right), \,\, \eta_k = \mathrm{True}\\
    0, \,\, \text{otherwise}
    \end{cases}
\end{align}
where the derivative propagates through the user representation to its component item embeddings as follows:
\begin{align}
    \frac{\partial \vec{u}_i}{\partial \vec{x}_j} = \frac{\vec{x}_j^2\gamma_{\vec{x}_j}^3 + \gamma_{\vec{x}_j}}{\sum_k \gamma_{\vec{x}_k}} - \frac{\vec{x_j}\sum_{k}\vec{x}_k\gamma_{\vec{x}_k}}
    {\left(\sum_k \gamma_{\vec{x}_k}\right)^2}
\end{align}
\end{document}